# PENERAPAN TEKNOLOGI PENGOLAH CITRA DIGITAL DAN KOMPUTASI PADA PENGUKURAN DAN PENGUJIAN BERBAGAI PARAMETER KAIN NON WOVEN

Andrian Wijayono[1], Irwan[1] & Valentinus Galih Vidia Putra[1]

Textile Engineering Departement, Politeknik STTT Bandung, Indonesia[1]

**Abstrak:** Salah satu tantangan besar industri saat ini adalah mengenai cara untuk memproduksi hasil produksi yang berkualitas, salah satunya adalah pada industri non woven. Peningkatan proses evaluasi dan pengendalian mutu produksi non woven telah banyak dikembangkan untuk mendukung peningkatan kualitas hasil produksi. Pemanfaatan teknologi informasi dan komputasi saat ini telah banyak diterapkan pada proses pengendalian mutu produksi bahan tekstil, salah satunya adalah pemanfaatan teknologi *image processing* pada proses evaluasi bahan non woven. Pada paper ini akan dijelaskan mengenai berbagai metoda penerapan teknologi *image processing* pada bidang evaluasi dan pengendalian mutu produksi tekstil.

**Kata Kunci:** *textile fiber, image processing, textile evaluation.*

# 1. PENDAHULUAN

Perkembangan teknik komputasi yang dinamis saat ini telah menciptakan kemungkinan besar untuk dapat diterapkan pada berbagai bidang, termasuk untuk mengindentifikasi dan mengukur dimensi geometris benda-benda yang sangat kecil termasuk benda tekstil. Dengan menggunakan analisis citra digital memungkinan dilakukannya analisis parameter struktural dasar benang tekstil yang lebih rinci, seperti ketebalan benang, *hairiness* dan jumlah *twist*. Teknik ini juga memungkinkan dilakukannya pengukuran dan identifikasi terhadap berbagai macam sifat dari struktur pada benang, seperti nomor benang [?], pengaruh pemberian antihan [?], dll. Proses identifikasi parameter struktur benang merupakan masalah signifikan pada proses penelitian ilmiah dan praktik industri saat ini sampai sekarang. Berdasarkan berbagai macam tinjauan literatur, dapat dikatakan bahwa teknik pemrosesan citra digital dapat dilakukan pada citra penampang melintang dari serat, diameter serat, serta kemungkinan untuk menganalisis citra benang untuk mengetahui jenis cacat dan penyebab cacat yang ada pada benang.

# 2. ANALISIS DAN PENGUKURAN KEKASARAN PERMUKAAN KAIN NONWOVEN DENGAN MENGGUNAKAN METODE MACHINE VISION

Teknik analisis citra digunakan dalam banyak aspek *engineering*, teknik ini juga telah digunakan untuk mengukur dan mengindentifikasi bahan tekstil dan *non woven*. Khususnya pada kain tenun, analisis citra adalah teknik yang cukup andal dan tepercaya untuk mengukur keseragaman, *cover factor*, kekasaran permukaan (surface *roughness)*, dan lain-lain. Beberapa topik penelitian telah dilakukan untuk mengukur keseragaman massa kain *non woven,* karena keseragaman berat kain merupakan faklor yang penting dalam struktur berserat tersebut (Cahhabra (2003) dan Militky (2007)). Ada pula penelitian yang dilakukan untuk menganalisa struktur kain *non woven* dengan menggunakan analisis citra (Pourdeyhimi dkk (2006) dan Huang dkk (1993)). Pada penelitian yang dilakukan oleh Semnani, Yekrang dan Ghayoor (2009), kain *non woven* dimodelkan menjadi struktur 3 dimensi untuk mengukur kekasarannya; Sebenarnya pemodelan lapisan tekstil berformat 3-D untuk mengukur sifat mereka bukanlah ide baru dan juga digunakan pada penelitian lain (Sul dkk(2006), Pourdyhimi dkk (2007) dan Hu dkk (2002)). Dalam Semnani dkk (2009), kain *non woven* disimulasikan dalam struktur virtual 3-D dan model geometris, struktur tersebut dianggap mensimulasikan lapisan *web* kain *non woven*. Pada metoda lain, tekstil dimodelkan dengan bentuk 3 dimensi untuk mengukur kerut kain dengan teknik stereo fotometrik. Selain itu, pada penelitian yang dilakukan oleh Sul dkk (2006), kekasaran permukaan diukur menggunakan data profil tiga dimensi. Dalam sistem tersebut, Sul dkk menggunakan metode penghitungan kotak untuk menghitung dimensi fraktal dan kemudian merekonstruksi citra untuk mendapatkan data kekasaran permukaan. Selain itu, beberapa peneliti mempresentasikan metode baru pada pengukuran objektif kekasaran permukaan kain berdasarkan analisis citra digital. Pada penelitian yang dilakukan oleh Sul dkk (2006), telah diketahui dua parameter pengukuran yang dapat dilakukan pada citra digital kain yang berdimensi fractal, yaitu dapat dihitung dengan *wavelet fractal method* dan *average mean curvature*. Kedua parameter ini dapat menggambarkan kekasaran permukaan (Kang dkk (2005) dan Kim dkk (2005)). Dalam penelitian lain, telah ditemukan pula metode untuk mengukur kekasaran permukaan dari suatu kain (Fontaine dkk, 2015).

Sistem pengukuran kekasaran permukaan konvensional sistem Kawabata. Pada sistem Kawabata, sebuah kawat baja berdiameter 0,5 mm dan berbentuk U ditarik dengan gaya sebesar 10 gf. Ketika kawat ditark dan bergeser, maka kawat akan bergerak bergeser sambil bergerak naik dan turun mengikuti kontur kain. Pergerakan kawat naik dan turun tersebut yang diukur sebagai kekasaran kain.

Ada dua permasalahan yang dihadapi pada pengukuran kain dengan metode Kawabata. Pertama, kekasaran kain tidak dapat diartikan sebagai perubahan ketinggian kontur pada kain. Kedua, friksi pada kain bukanlah hal yang yang sangat penting pada pengukuran kekasaran kain. Koefisien friksi pada kain sebenarnya dianggap sebagai suatu kriteria dari pegangan kain, dan bukan sebagai kekasaran permukaan.

Pada penelitian yang dilakukan oleh Govindaraj dkk (2003), telah ditemukan suatu metode untuk mensimulasikan kekasaran kain sebagai sebuah fungsi sinyal yang direfleksikan oleh kain. Pada metoda tersebut telah berhasil dihasilkan suatu metoda untuk merekonstruksi bentuk permukaan dari suatu kain (Govindaraj dkk, 2003). Penggunaan metoda ini masih dibatasi penggunaannya hanya untuk kain tenun. Kain dengan struktur arah serat yang acak (random) tidak dapat dianalisis dengan menggunakan sistem tersebut.

Pada penelitian yang dilakukan oleh Semnani, Yekrang dan Ghayoor (2009), telah ditemukan suatu metode untuk memodelkan struktur permukaan lapisan kain *nonwoven* dengan mengubah citra 2-D kain ke dalam bentuk struktur 3-D. Pada penelitian tersebut juga telah ditemukan suatu hubungan antara struktur kain dengan friksi kain.

Untuk mengevaluasi kekasaran permukaan dibutuhkan suatu acuan pembanding agar permukaan kain bisa dibandingkan. Sebenarnya acuan ini tidak lain adalah lapisan yang menghadirkan kekasaran yang paling memberikan rasa nyaman untuk sensasi sentuhan manusia, lapisan ini disebut permukaan ideal.

Permukaan yang ideal adalah permukaan gelombang sinusoidal dengan amplitudo dan panjang gelombang yang dapat dirasakan oleh sensasi sentuhan manusia, permukaan ini memiliki rasa pegangan yang paling nyaman untuk tekstil. Permukaan yang ideal juga dapat menimbulkan gesekan ideal tekstil untuk tubuh manusia. Koefisien gesekan tekstil umumnya terkait dengan sifat sentuhan dan pegangannya (Bertaux dkk (2007) dan Sular (2008)), kain terbaik dari sudut pandang kekasaran dianggap sebagai kain yang memiliki tinggi gelombang sesuai dengan sentuhan manusia.

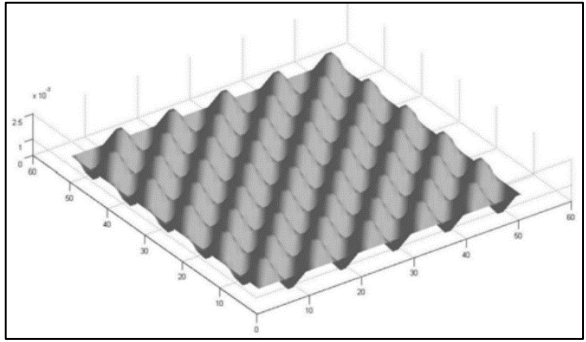

Gambar-1 Simulasi pemodelan kain ideal

Tentunya sentuhan jari sensitif terhadap gesekan, karena itu model disimulasikan berdasarkan jaringan jari manusia. Menurut penelitian Semnani (2009), panjang gelombang dan amplitudo yang tidak akan dirasakan oleh manusia adalah sebesar 1 mm dan 2,5 µm. Gambar-1 menggambarkan simulasi permukaan yang ideal sesuai dengan data yang disebutkan. Dimana bidang x-y adalah jumlah matriks pixel dan arah z adalah tinggi.

Validasi eksperimen dilakukan dengan menggunakan kain contoh uji *nonwoven* yang dihasilkan dari serat polypropylene 0,87 dtex. Contoh uji dipindai citranya pada resolusi 600 DPI menggunakan perangkat scanner. Latar contoh uji yang digunakan adalah warna hitam. Citra yang diperoleh dikonversi kedalam citra *grey scale* pada skala 256 keabuan, setelah itu digunakan filter Wiener dan Gaussian untuk mengurangi jumlah noise pada gambar. Pada gambar tersebut, daerah yang terang menunjukan bagian yang tebal dari lapisan non woven dan daerah yang gelap menunjukan bagian yang relatif tipis pada lapisan *nonwoven*. Gambar-2 menunjukan hasil pindaian citra kain contoh uji *nonwoven*, sedangkan Gambar-2(b) menunjukan gambar yang telah diproses menggunakan metode *histogram equalization* agar didapatkan citra digital yang lebih baik.

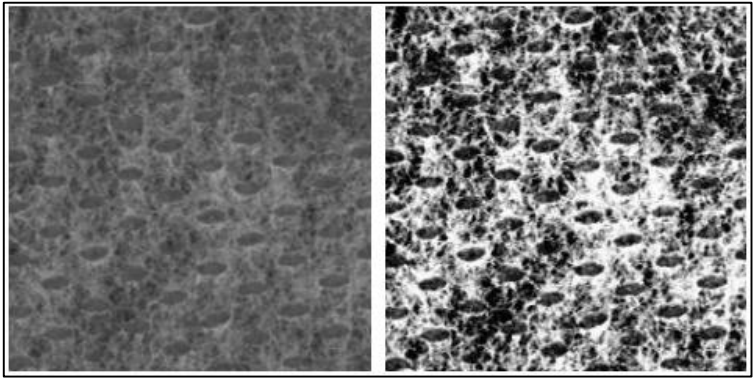

(a) (b)

Gambar-2 Citra hasil pemindaian dari kain *nonwoven*

Kemudian setiap pixel dari citra pada Gambar-2(b) dipetakan pada grafik simulasi profil permukaan seperti pada Gambar-3. Profil permukaan telah digambar dengan dasar stiap elemen dalam profil ini disimulasikan tinggi pada titik gambar. Daerah paling terang pada citra dipetakan sebagai titik tertinggi pada profil tersebut, maka *pixel* yang memiliki skala abu 255 (putih) akan digambarkan sebagai titik tertinggi pada lapisan yang tingginya 2,5 µm. Setelah diperoleh profil tersebut, maka dapat dengan mudah untuk dapat dinilai kekasaran permukaannya, yaitu dengan cara membandingkannya dengan model simulasi yang telah ditetapkan sebelumnya.

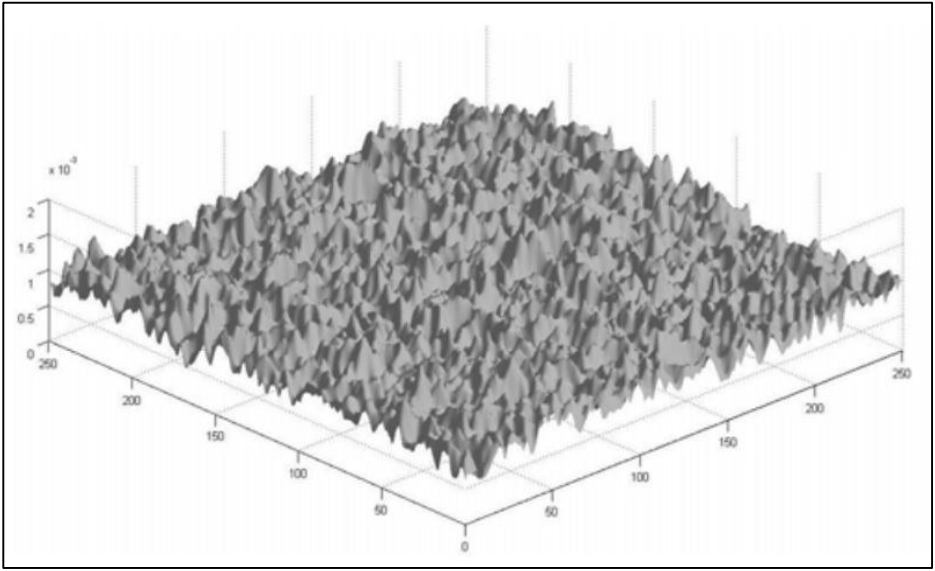

Gambar-3 Simulasi profil dari kain *nonwoven* contoh uji

Pada penilaian profil kain *nonwoven*, ada lima kriteria dari profil tersebut yang akan dievaluasi, diantaranya (1) $N$, yaitu jumlah puncak pada permukaan nonwoven; (2) $T$, yaitu variasi jarak antar puncak pada profil *nonwoven*; (3) $E$, yaitu volume kain yang didasarkan pada profl yang telah diambil pada kain; (4) $I_d$, yaitu variasi selisih nilai gray scale terhadap nilai rata-rata gray scale; dan (5) $V$, yaitu variansi dari nilai *gray scale* tiap puncak pada profil. Kelima kriteria ini membantu kita untu menentukan suatu faktor yang kita sebut sebagai *surface roughness* ($R_s$), menggunakan kelima kriteria ini kita dapat mengevaluasi faktor *surface roughness.* Nilai dari setiap elemen (*pixel*) sebenarnya adalah merepresentasikan tingginya nilai yang disimulasikan dan digunakan untuk menilai kekasaran atau *roughness*. Ide penggunaan ketinggian nilai untuk mengukur *roughness* sebenarnya datang dari sistem evaluasi Kawabata. Adapun kelemahan yang telah diperbaiki dari sistem Kawabata, yaitu pengukuran baru ini bersifat *point to point* (penilaian dilakukan pada setiap titik pada kain), berbeda dengan sistem Kawabata yang mengukur dengan sistem *line to line* (perabaan ketinggian menggunakan kawat, sehingga tidak dapat menampilkan hasil yang valid (Semnani dkk, 2009).

Untuk mengevaluasi metoda Semnani dkk (2009), telah dilakukan pengujian terhadap 15 buah contoh uji kain *nonwoven* yang dibandingkan koefisien friksinya. 15 kain tersebut telah diukur koefisien friksinya berdasarkan metode ASTM D1894 dan *surface roughness*-nya berdasarkan metode Semnani (2009). Perbandingan nilai antara hasil pengujian kedua metode pengujian tersebut dapat dilihat pada Tabel-1.

Mesin Zwick *testing* telah digunakan untuk mengukur gaya friksi selama proses pergeseran kain. Pengukuran gaya tersebut menghasilkan nilai koefisien friksi, yang dihitung dari hasil pembagian nilai rata-rata gaya friksi terhadap nilai gaya normal sebesar 28 cN.

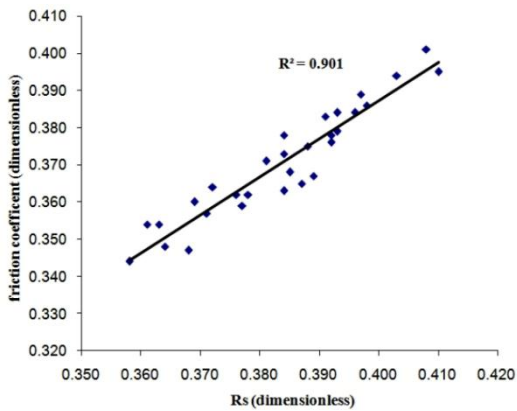

Gambar-4 Hubungan antara koefisien friksi dan $R_s$

Tabel-1 Hasil pengukuran *surface roughness* dan koefisien friksi kain

| Sampel | Surface Roughness | Friction Coefficient |
|--------|-------------------|----------------------|
| 1      | 0.410             | 0.395                |
| 2      | 0.376             | 0.362                |
| 3 | 0.364 | 0.348 |
| 4 | 0.368 | 0.347 |
| 5 | 0.384 | 0.373 |
| 6 | 0.398 | 0.386 |
| 7 | 0.384 | 0.378 |
| 8 | 0.378 | 0.362 |
| 9 | 0.371 | 0.357 |
| 10 | 0.403 | 0.394 |
| 11 | 0.369 | 0.360 |
| 12 | 0.361 | 0.354 |
| 13 | 0.408 | 0.401 |
| 14 | 0.389 | 0.367 |
| 15 | 0.391 | 0.383 |
| 16 | 0.381 | 0.371 |
| 17 | 0.384 | 0.363 |
| 18 | 0.392 | 0.376 |
| 19 | 0.388 | 0.375 |
| 20 | 0.387 | 0.365 |
| 21 | 0.392 | 0.378 |
| 22 | 0.377 | 0.359 |
| 23 | 0.393 | 0.379 |
| 24 | 0.396 | 0.384 |
| 25 | 0.385 | 0.368 |
| 26 | 0.397 | 0.389 |
| 27 | 0.372 | 0.364 |
| 28 | 0.358 | 0.344 |
| 29 | 0.393 | 0.384 |
| 30 | 0.363 | 0.354 |

Gambar-4 menunjukan perbandingan antara $R_s$ dan koefisien friksi. Telah ditemukan pula hasil korelasi yang baik antara $R_s$ dan koefisien friksi. Koefisien regresi antara koefisien friksi dan $R_s$ adalah 0,901 dan dapat diyakinkan bahwa hubungannya adalah berbanding lurus.

Berdasarkan hal tersebut, maka image processing juga dapat diaplikasikan pada proses evaluasi *surface roughness* kain dan dapat menghasilkan nilai yang bersifat valid melalui pengukuran komputasi. Semakin besar *surface roughness* atau kekasaran permukaan suatu kain, maka akan semakin besar juga koefisien friksinya, kedua variabel tersebut memiliki hubungan $\mu = 1,027\ R_s - 0,023$ .

## 3. ANALISIS MIKROSTRUKTURAL UNTUK KAIN NONWOVEN MENGGUNAKAN PEMINDAIAN MIKROSKOP ELEKTRON DAN PENGOLAH CITRA

Teknik pengolah gambar banyak digunakan untuk mengevaluasi parameter mikrostruktur kain *nonwoven* seperti distribusi orientasi, distribusi diameter, kerutan, kain tanpa keseragaman dan cacat. Aspek mikrostruktur terpenting dari kain bukan tenunan yang berkaitan dengan serat yang digunakan dalam produksi kain adalah diameter, panjang, bentuk kerutan dan penampang melintang. Parameter geometris pada skala mikro yang berkaitan dengan struktur kain adalah distribusi orientasi, distribusi curl dan jumlah titik kontak.

Berbagai teknik digunakan untuk memperkirakan distribusi orientasi pada kain *nonwoven*, seperti *direct tracking*, analisis *flowfield* dan metode transformasi Fourier (FT) dan metode Hough Transform (HT). Dalam *direct tracking*, garis *pixel* yang sebenarnya akan dilacak (Pourdeyhimi dan Ramanathan, 1996). Algoritma yang digunakan dapat memakan waktu dan oleh karena itu banyak dilakukan penelitian dan pengembangan lebih lanjut. Pada struktur yang lebih padat, karena jumlah titik silang meningkat, metode ini menjadi kurang efisien untuk digunakan.

Analisis *flowfield* didasarkan pada asumsi bahwa tepi pada gambar mewakili bidang orientasi pada gambar. Metode ini terutama digunakan untuk mendapatkan sudut distribusi orientasi rata-rata, namun bukan fungsi distribusi orientasi (Pourdeyhimi dan Ramanathan, 1997).

Metode FT banyak digunakan dalam banyak pengolahan citra dan operasi pengukuran. Ini mengubah domain intensitas skala abu-abu menjadi spektrum frekuensi (Pourdeyhimi dan Ramanathan, 1997). Kompresi gambar, penyamaratan kecerahan gambar dan kontras dan penyempurnaan dapat lebih mudah dilakukan pada spektrum frekuensi. Dalam kebanyakan kasus domain spasial perlu dipulihkan dari spektrum frekuensi. Mengukur orientasi adalah salah satu aplikasi metode FT. Algoritma ini cepat dan tidak memerlukan daya komputasi yang tinggi untuk perhitungannya.

Metode HT digunakan untuk memperkirakan distribusi orientasi serat secara langsung (Xu dan Yu, 1997). Keuntungan dari metode ini dibandingkan dengan metode tidak langsung lainnya adalah bahwa orientasi sebenarnya dari garis tersebut termasuk secara langsung dalam perhitungan transformasinya. Kemampuan metode HT dalam pengenalan objek digunakan untuk mengukur panjang garis lurus pada gambar. Namun, diperlukan waktu dan sumber komputasi yang lebih banyak untuk menjalankan algoritma untuk gambar berskala lebih besar dengan akurasi yang lebih tinggi.

Pourdeyhimi dkk (1996, 1997) pertama kali menguji metode ini pada gambar simulasi *web nonwoven* dan kemudian menerapkannya pada gambar jaring sebenarnya. Mereka melaporkan sejumlah masalah, sebagian besar berkaitan dengan perolehan gambar, dalam memperluas metode ke *web* yang sebenarnya. Mereka bereksperimen dengan sejumlah sistem pencahayaan, termasuk optik *back lighting* dan berbagai mikroskop optik dengan medan gelap - serta kemampuan medan terang dan berbagai filter *polarizing*. Gambar yang diperoleh dengan menggunakan backlighting dan mikroskop optik dilaporkan memiliki kualitas rendah dengan kontras yang buruk. Oleh karena itu, Pourdeyhimi dkk (1996, 1997) membatasi teknik akuisisi gambar ke jaring yang sangat tipis (ringan) sehingga mereka bisa mendapatkan gambar yang bisa berhasil diolah (Pourdeyhimi, 1999). Untuk mengatasi masalah ini, pada penelitian ini dilakukan mikroskop elektron scanning (SEM) yang telah digunakan untuk menghasilkan gambar berkualitas tinggi. SEM menyediakan skala pembesaran yang tinggi. Teknik ini tidak terbatas pada jaring tipis dan juga bisa digunakan untuk kain dengan kerapatan yang sangat tinggi.

Dalam penelitian ini, kemampuan metode transformasi *fast Fourier Transform* (FFT) dan HT untuk mengevaluasi distribusi orientasi serat dalam sampel gambar tertentu telah dievaluasi. Efek curl pada panjang perkiraan juga telah diselidiki. Efek dari *preprocessing* dan parameter akuisisi citra yang berbeda pada hasil pengolahan citra telah dipelajari.

Pada penelitian yang telah dilakukan oleh Ghassemieh dkk (2002), citra kain *nonwoven* hasil pemindalan dianalisis dengan *fourier transformation,* sehingga citra berubah dalam bentuk *discrete fourier transform* (seperti dapat dilihat pada Gambar-4b).

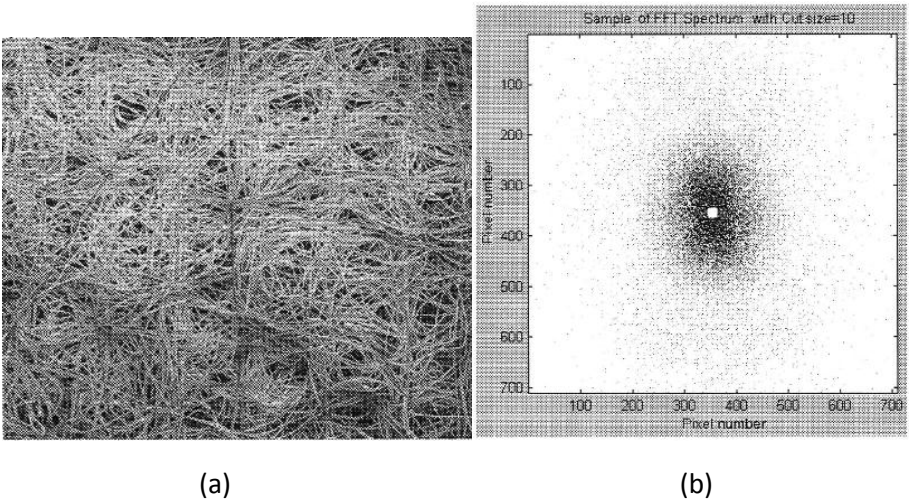

(a) (b)

Gambar-4 Citra digital hasil tangkapan mirkoskop SEM (a) dan hasil analisis citra menggunakan analisis FT

Analisis FT dua dimensi menguraikan suatu citra yang merupakan domain spasial yang terdiri dari intensitas-intensitas warna dan mengubahnya dapat bentuk domain frekuensi. Pada domain frekuensi, perubahan pada posisi gambar bergantung pada perubahan pada frekuensi spasial atau rerata intensitas gambar yang berubah pada domain spasial citra $I$, sehingga nilai $F$ yang merepresentasikan jumlah nilai intensitas pada gambar $I$ dapat ditulis sebagai berikut

$$F(u,v) = \int_{-\infty}^{\infty}\int_{-\infty}^{\infty} f(x,y)\, e^{-2\pi i(xu+yv)} dxdy \qquad (1)$$

Setelah gambar ditangkap, untuk melakukan analisis FT, maka rumusan dari FT yang dibutuhkan adalah sebagai berikut

$$F(u,v) = \frac{1}{N} \sum_{x=0}^{N-1} \sum_{y=0}^{N-1} f(x,y) e^{-2\pi i(xu+yv)/N} \qquad (2)$$

Persamaan tersebut juga dapat dituliskan sebagai berikut

$$F(u,v) = \frac{1}{N} \sum_{x=0}^{N-1} e^{-2\pi i xu/N} \sum_{y=0}^{N-1} f(x,y) e^{-2\pi i yv/N} \qquad (3)$$

Penjumlahan pertama pada dasarnya adalah DFT satu dimensi jika x memiliki suatu nilai. Penjumlahan yang kedua adalah DFT satu dimensi yang dilakukan dengan hasil penjumlahan yang pertama. DFT dua dimensi dapat dihitung dengan melakukan DFT satu dimensi untuk setiap nilai dari $x$, misal untuk setiap kolom dari $f(x,y)$, kemudian lakukan pengukuran DFT satu dimensi pada arah yang berlawanan (untuk tiap baris). Algoritma tersebut kemudian disebut sebagai *fast fourier transform*, yang dapat mengurangi jumlah transformasi sebanyak $N^2$ menjadi $N \log^2 N$ dan membuat proses komputasi pada komputer menjadi lebih cepat untuk dilakukan.

Transformasi FT sangat berguna untuk menentukan rata-rata intensitas transisi yang terjadi pada sebuah arah yang diberikan pada suatu citra digital. Berdasarkan hal tersebut, misalkan apabila hampir keseluruhan serat sejajar dengan arah kain *nonwoven*, maka nilai frekuensi spasial pada arah tersebut akan kecil, sedangkan nilai frekuensi spasial pada arah tegak lurusnya akan besar. Analisis FT tersebut telah digunakan untuk memperoleh orientasi serat pada suatu kain *nonwoven.*

Transformasi diimplementasikan dengan memproses seluruh baris pada waktu yang bersamaan, diikuti oleh seluruh kolom pada waktu yang sama. Hasil dari transformasi tersebut adalah sebuah kumpulan nilai yang tersusun pada matriks dua dimensi yaaang memiliki suatu besaran dan sebuah fase. Besarnya dari frekuensi tersebut diindikasikan dengan intensitasi *pixel* pada lokasi tersebut (dapat dilihat pada Gambar-4b). Daerah yang lebih gelap menunjukan nilai yang lebih besar. Komponen frekuensi yang bernilai nol sesuai terhadap nilai rata-rata kecerahan citra digital.

Selain metode FT, dikenal pula metode HT (Hough transform), adalah salah satu metode yang sangat efektif untuk mendeteksi objek pada suatu citra digital. Umumnya, analisis metode HT digunakan untuk mendeteksi garis yang menempati suatu bangun seperti busur, lingkaran, elips, dll. Pada penelitian yang dilakukan oleh Gassemieh dkk (2002) adalah untuk mendeteksi bagian lurus dari serat pada kain *nonwoven.*

Untuk mendefinisikan garis lurus dibutuhkan dua parameter. Pada koordinat kartesian, persamaan garis lurus dapat ditulis sebagai berikut

$$y = mx + b \qquad (4)$$

$m$ merupakan slope dan $b$ adalah nilai *intercept*. Oleh karena $m$ bersifat infinit untuk garis yang parallel terhadap sumbu $y$, maka representasi ini tidak dapat digunakan secara baik dalam analisis HT. Sistem koordinat polar yang digunakan dalam merepresentasikan garis yang digunakan adalah sebagai berikut

$$\rho = x \cos\theta + y \sin\theta \qquad (4)$$

$\rho$ adalah jarak titik tegak lurus dari titik awal dan $\theta$ adalah sudut yang terbentuk antara garis dengan sumbu normal. Gambar-5a menunjukan skema suatu garis yang terdiri atas jejari $\rho$ dan sudut $\theta$. Hal tersebut memungkinkan sudut bervariasi antara 0 hingga $2\pi$ dan $\rho$ akan selalu bernilai positif, atau dapat pula dimungkinkan untuk $\rho$ bernilai positif atau negative dengan cara membatasi nilai sudut $\theta$ pada interval 0 hingga $\pi$.

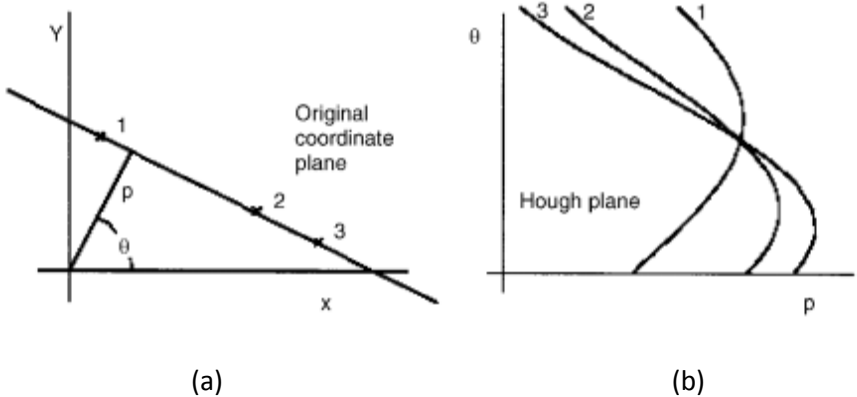

(a) (b)

Gambar-5 Contoh dari garis pada sistem koordinat kartesian (a) dan spectrum Hough dari contoh uji (b)

Titik collinear $(x_i, y_i)$ dengan $i = 1, ..., N$ adalah transformasi pada bentuk kurva sinusoidal $\rho = x \cos \theta + y \sin \theta$ pada bidang Hough, yang akan berpotongan pada titik $(\rho, \theta)$(dapat dilihat pada Gambar-5b). Ketika nilai biner $(\Delta\rho, \Delta\theta)$ pada bidang Hough terlalu halus, setiap perpotongan dari kurva sinusoidal dapat memiliki biner yang berbeda. Ketika proses kuantisasi tidak cukup halus, di lain pihak, garis sejajar tersebut terlalu dekat akan jatuh pada titik bin yang sama.

Untuk interval tertentu dari nilai yang kuantisasi dari parameter $\rho$ dan $\theta$, tiap $(x_i, y_i)$ dipetakan pada bidang Hough dan titik-titik pada peta ditempatkan pada $(\rho_m, \theta_m)$ diakumulasikan pada histogram dua dimensi. Titik-titik maksmimum pada bidang Hough diidentifikasi sebagai bagian garis lurus pada bidang citra sebenarnya.

Ketika metode diaplikasikan pada kain *nonwoven*, pertama garis *pixel* dari serat akan dideteksi dengan menggunaka *edge detection* (pendeteksi pinggiran) dan seluruh *pixel* pada citra yang memiliki nilai lebih besar dari nilai ambang batas *thresholding* didefinisikan sebagai *pixel serat.* Proses tesebut akan menghasilkan sebuah citra ciner yang dapat dilihat pada Gamba-6a.

Dengan menggunakan analisis Hough, maka dapat didapatkan grafik histogram pada Gambar-6b.

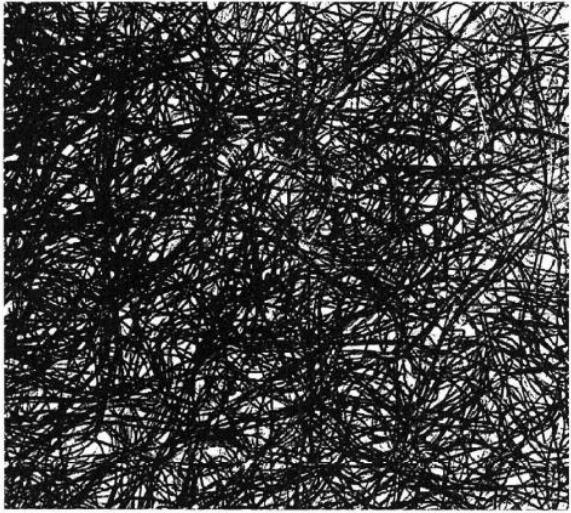

(a)

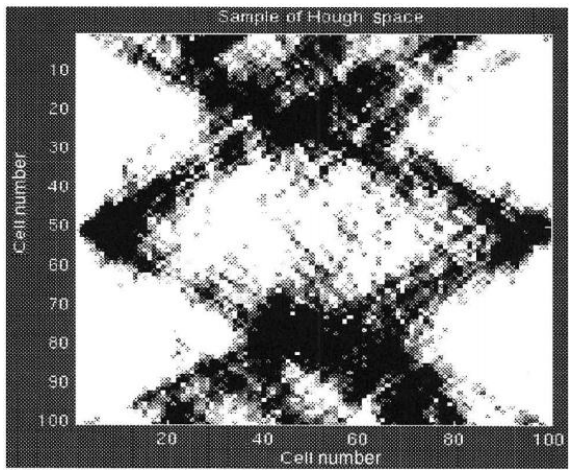

(b)

Gambar-6 Citra biner kain *nonwoven* (a) dan grafik histogram analisis HT (b)

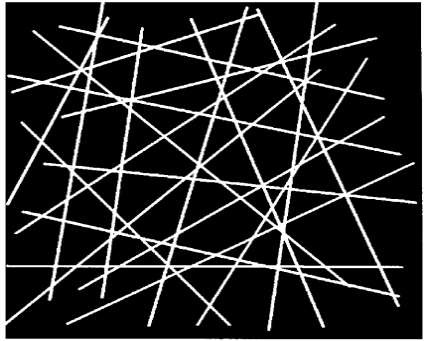

(a)

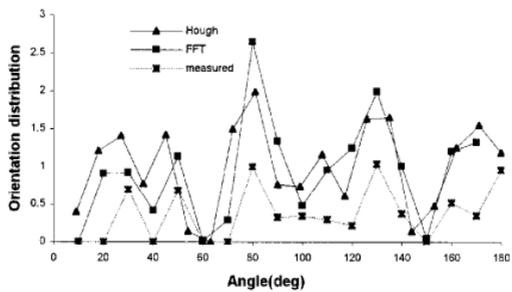

(b)

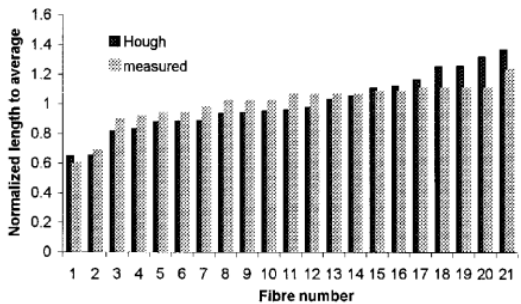

(c)

Gambar-7 Hasil test contoh garis dari kain nonwoven (a), perbandingan orientasi serat antara hasil analisis FT dan analisis HT (b) dan perbandingan jumlah serat untuk tiap nilai sudut berdasarkan analisis FT dan HT

Untuk memvalidasi teknik yang dikembangkan dan menentukan akurasi dalam evaluasi orientasi serat dan panjang distribusi serat, telah dilakukan beberapa pengujian sampel terdiri dari beberapa garis yang merepresentasikan serat pada sebuah kain *nonwoven*. Gambar-7a menunjukan suatu ilustasi dari garis-garis yang merepresentasikan orientasi serat pada konstruksi *nonwoven*. Setelah dilakukan analisis HT dan FT pada kain *nonwoven*, telah didapatkan hasil sudut orientasi serat seperti pada grafik Gambar-7b dan Gambar-7c. Berdasarkan analisis pada grafik Gambar-7b, didapatkan bahwa analisis HT mampu memberikan hasil dengan simpangan terhadap rata-rata nilai orientasi yang lebih baik.

Gambar-8 dan Gambar-9 menunjukan perbandingan antara hasil analisis orientasi serat dengan metode representasi serat sebagai garis lurus dan metode representasi serat sebagai garis lengkung. Dapat dilihat pada grafik Gambar-9 bahwa dengan representasi garis lengkung, distribusi orientasi hasil analisis FFT dan HT memiliki tidak berkorelasi dengan representasi garis lurus. Analisis HT tersebut merupakan proses diskrit dari nilai-nilai yang ada pada citra untuk dibuat nilai dalam suatu bidang HT. Panjang distribusi serat (dengan metode HT) dari garis lengkung akan menghasilkan panjang serat yang lebih pendek dibandingkan dengan representasi garis lurus. Nilai simpangan dari rata-rata panjang serat juga berkurang secara signifikan untuk metode lengkung, seperti yang dapat dilihat pada Gambar-9c.

Hasil pengujian menunjukan bahwa FFT, HT dan metode pengolahan *preprocessing* citra dapat mengidentifikasi secara benar distribusi orientasi serat. Hasil analisis HT dari sampel uji membuktikan validitas dari metode estimasi panjang garis serat.

Adapun beberapa parameter yang dapat mempengaruhi hasil analisis seperti pembesaran pada gambar, kerataan kecerahan pada gambar, pengaruh perlakukan *preprocessing* dan bentuk *frame* citra kain *nonwoven*. Pengaruh perbesaran citra pada gambar kain *nonwoven* dapat dilihat pada Gambar-10. Pengaruh bentuk *frame* gambar kain *nonwoven* dapat dilihat pada Gambar-11.

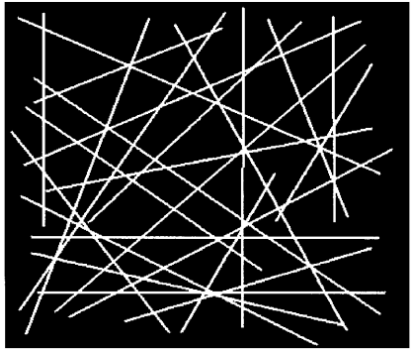

(a)

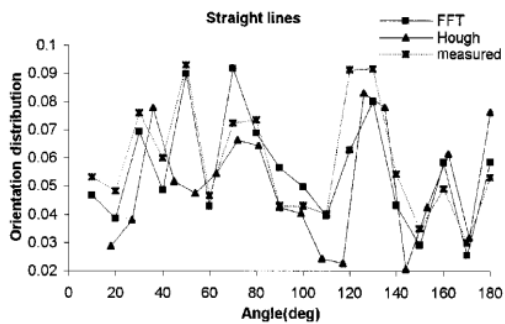

(b)

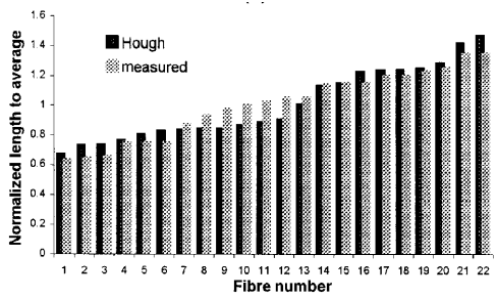

(c)

Gambar-8 Sampel pengujian garis (a), hasil analisis dengan metode FFT dan HT (b), dan perbandingan panjang serat dari hasil pengukuran dan hasil analisis HT

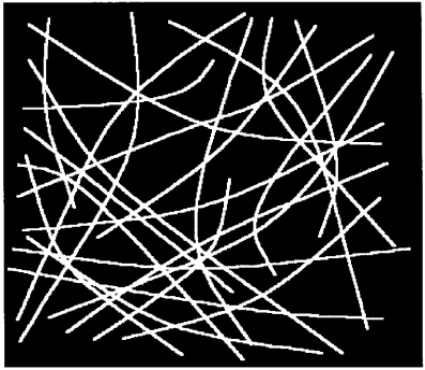
(a)

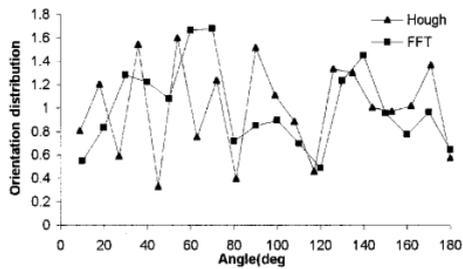
(b)

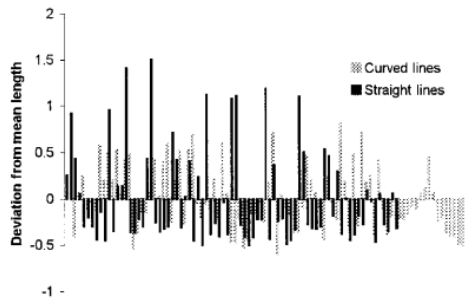
(c)

Gambar-9 Sampel hasil pengujian garis lengkung (a), hasil analisis dengan metode FFT dan HT (b), dan perbandingan panjang serat dari hasil pengukuran garis lurus dan garis lengkung

Perbesaran dan besar daerah pada gambar merupakan faktor yang penting dan dapat mempengaruhi hasil analisis. Parameter ini harus dioptimasi agar didapatkan suatu gambar yang representative dari keseluruhan kain dan dapat dikenali oleh metode pengolahan citra digital. Pada penelitian yang dilakukan oleh Ghassemieh dkk (2002), perbesaran yang telah dilakukan pada citra SEM adalah sebesar 30x, 50x, dan 100x. Hasil distribusi orientasi serat dengan menggunakan FFT dapat dilihat pada Gambar-9. Dapat dilihat bahwa hasil analisis citra pada perbesaran 30x dan 50x memberikan hasil distribusi yang tidak terlalu jauh berbeda, sedangkan pada perbesaran citra 100x memberikan hasil yang benar-benar berbeda dengan perbesaran yang lainnya.

Faktor kedua yang dapat mempengaruhi hasil analisis distribusi serat adalah bentuk *frame* dari citra kain *nonwoven*. Bentuk *frame* dari citra mempengaruhi hasil dari analisis pengolahan citra FFT. Citra dengan bentuk *frame* persegi atau lingkaran telah dianalisis dengan metode FFT, hasilnya dapat dilihat pada Gambar-10. *Frame* persegi menunjukan nilai persentasi yang lebih kecil dari orientasi serat pada citra lingkaran. Pada citra dalam bentuk lingkaran, hal tersebut sedikit berbeda karena penentuan garis normal pada kain akan lebih sulit untuk dilakukan pada FFT.

Faktor ketiga yaitu pengaruh kerataan kecerahan gambar citra kain *nonwoven.* Kecerahan yang berbeda pada citra hasil tangkapan mirkoskop SEM dapat terjadi akibar distribusi elektron pada lembaran contoh uji tidak terjadi secara merata. Kecerahan citra yang tidak seragam tersebut akan mengakibatkan adanya perbedaan kontribusi dari bagian gambar terhadap hasil pengolahan citra. Citra SEM pada kain *nonwoven* yang kerataan cerahnya dapat dilihat pada Gambar-11a dan Gambar-11b. Hasil analisis terhadap nilai distribusi orientasi serat ditunjukan oleh Gambar-11c, yang menunjukan bahwa kecerahan yang tidak merata mempengaruhi hasil analisis secara signifikan. Oleh karena itu, perlu dipastikan bahwa cita SEM telah memiliki kecerahan yang merata. Pada metode HT, proses pengubahan citra digital menjadi citra biner perlu untuk dilakukan. Namun, ketidakrataan kecerahan pada gambar akan membuat sebagian serat pada citra akan hilang.

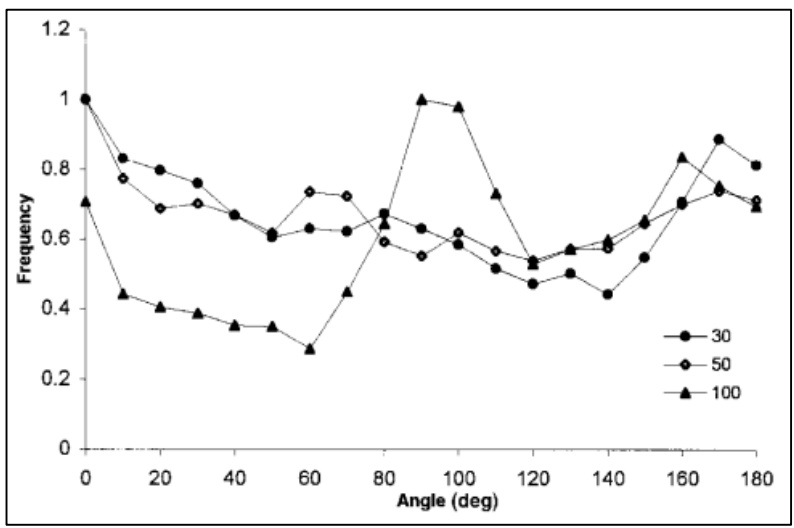

Gambar-9 Pengaruh pembesaran gambar dari citra kain *nonwoven* terhadap hasil analisis nilai distribusi orientasi serat

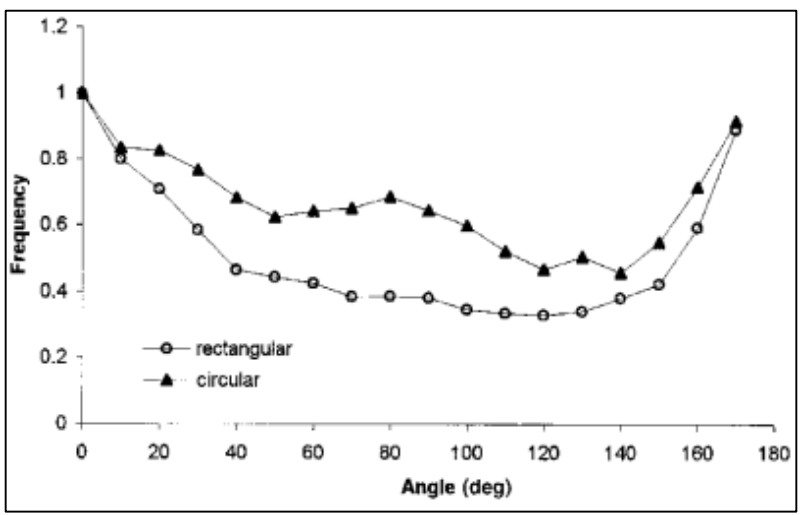

Gambar-10 Pengaruh bentuk *frame* dari citra kain *nonwoven* terhadap hasil analisis nilai distribusi orientasi serat

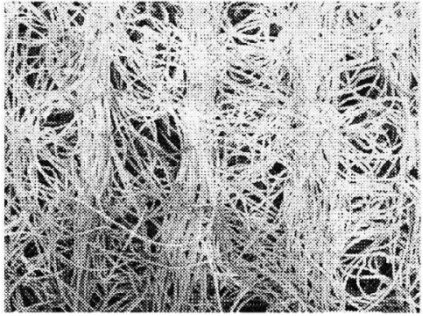

(a)

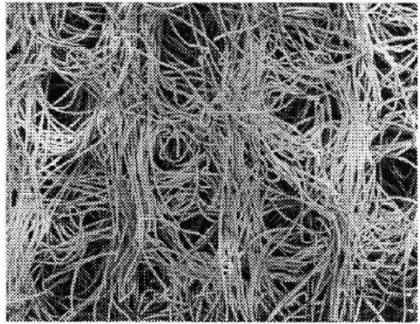

(b)

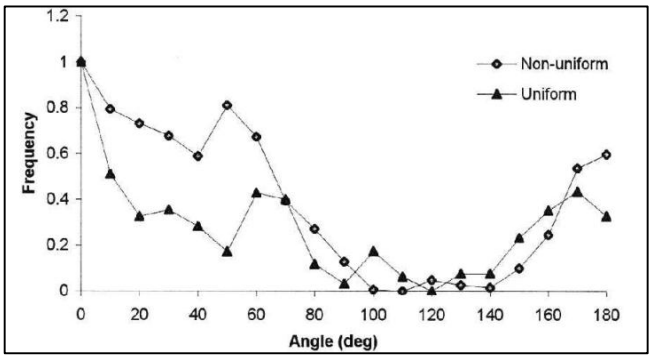

(c)

Gambar-11 Citra digital dengan kecerahan yang tidak merata (a), citra digital dengan kecerahan yang merata (b) dan perbandingan distribusi orientasi serat terhadap kerataan kecerahan citra (c)

Faktor keempat yang mempengaruhi hasil distribusi orientasi serat yaitu efek dari tahap *preprocessing* pada citra digital. Gambar-13 dan Gambar-14 merupakan citra *skeleton* dan biner dari citra kain *nonwoven* pada Gambar-12.

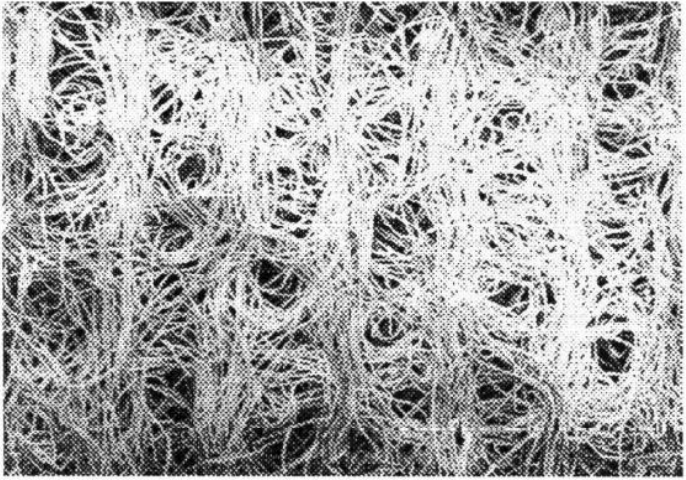

Gambar-12 Citra kain *nonwoven* yang ditangkap menggunakan mikroskop SEM

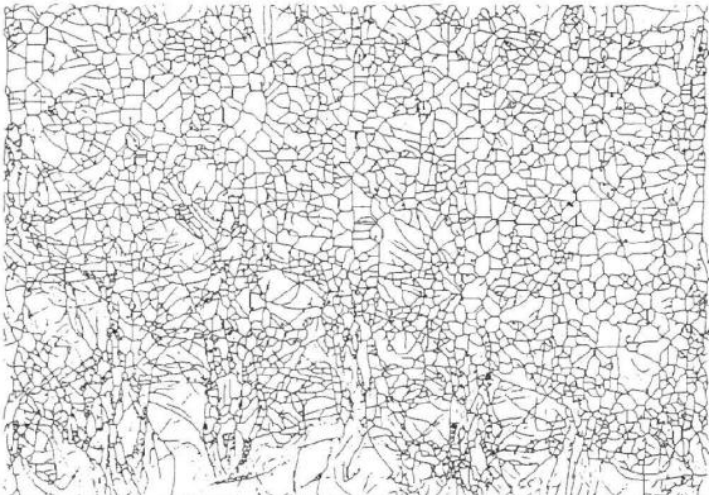

Gambar-13 Citra hasil proses *skeletoning process*

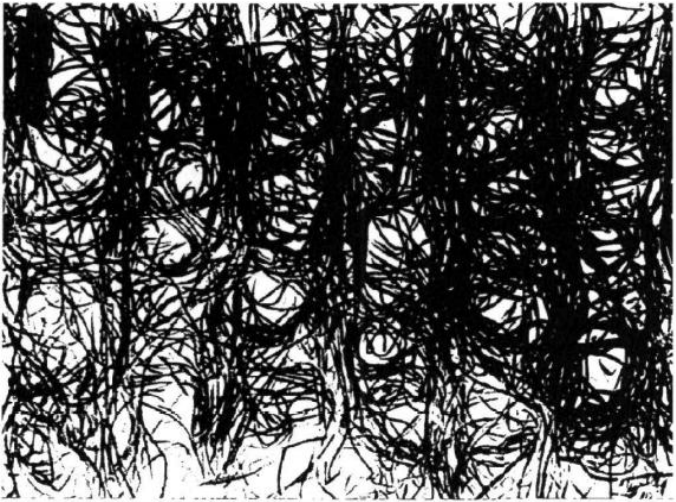

Gambar-14 Citra biner yang dihasilkan dari proses binerisasi

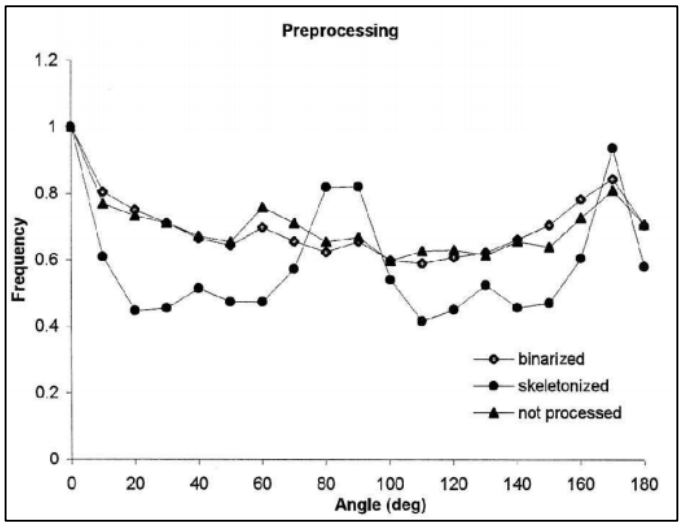

Gambar-15 Perbandingan distribusi orientasi serat

Berdasarkan hasil analisis distribusi orientasi serat yang telah dilakukan, dapat dikatakan bahwa dengan analisis FT dan HT dapat memberikan analisis citra digital suatu melalui proses komputasi. Telah dianalisis pula faktor-faktor yang dapat mempengaruhi hasil analisis distribusi orientasi serat. Dengan demikian diketahui bahwa *image processing* dapat diaplikasikan pada penentuan distribusi orientasi serat serat serta memberikan hasil yang akurat.

## 4. PENILAIAN OBJEKTIF PADA PARAMETER PILLING KAIN NONWOVEN MENGGUNAKAN TRANSFORMASI WAVELET DISKRIT DUA DIMENSI

Fabric pilling merupakan masalah serius bagi industri, yang menyebabkan penampilan kain tak bagus dipandang dan kerusakan dini pada kain. Produk wol sangat rentan terhadap pilling. Baru-baru ini, sebuah proses untuk memproduksi kain pakaian wol *nonwoven* telah dikomersialkan di Australia, dan dapat menumbuhkan pasar baru untuk wol Australia (Palmer, 2005). Kualitas kain *nonwoven* semacam itu sebagian akan bergantung pada karakteristik *pil* (gundukan serat pada permukaan kain). Elemen kunci dalam pengendalian *pilling* kain adalah evaluasi ketahanan terhadap *pilling* dengan pengujian *pilling* kain. Ketahanan terhadap *pilling* biasanya diuji di laboratorium dengan proses yang mensimulasikan gosokan pada kain, diikuti dengan penilaian manual tingkat *pilling* oleh ahli berdasarkan perbandingan visual sampel dengan satu set gambar uji standar. Untuk membawa lebih banyak objektivitas ke dalam proses *grading*, sejumlah sistem otomatis berdasarkan analisis citra telah dikembangkan.

*Pilling* adalah pembentukan serabut kecil dari serat atau bola pada permukaan kain yang dihasilkan akibat proses pencucian, pengujian atau pemakaian kain. 'Pil' pada permukaan kain membuat kain sangat tidak enak dipandang dan kain semacam itu ditolak oleh konsumen yang cerdas. *Pilling* telah menjadi masalah serius bagi industri pakaian jadi, yang secara tradisional didominasi oleh kain rajutan dan anyaman. Perkembangan pesat pakaian jadi dalam beberapa tahun terakhir telah menambahkan dimensi baru pada masalah dari pembuatan kain, dan tidak ada penelitian mendasar yang dilakukan pada pembuatan bahan kain *nonwoven*.

Kain *nonwoven* terdiri dari jaringan serat tipis yang tergabung menjadi satu. Proses *nonwoven* adalah proses mengubah serat-serat langsung menjadi kain yang relatif sederhana, dibandingkan dengan proses pemintalan serat menjadi benang yang digunakan untuk memproduksi kain tenun dan rajut. Bahan *nonwoven* berbeda dengan bahan tenun dan rajut dalam struktur dan kinerja. Tetapi kain *nonwoven* memiliki banyak aplikasi penting, yaitu sebagai bahan penyerap, tekstil medis, filter, geotekstil, produk serat alami, bahan komposit, tekstil kendaraan, bahan bangunan, bantalan, karpet dan isolasi. Aplikasi ini didominasi tekstil teknis yang diproduksi dari serat sintetis (David, 2003).

Australia menghasilkan wol kualitas terbaik, yaitu merino wool. Di seluruh dunia, lebih dari 70% wol untuk penggunaan pakaian berasal dari Australia. Total produksi wol pada tahun 2000-2001 senilai 2,5 juta dollar Australia, wol juga merupakan sekitar 7% dari hasil keseluruhan pertanian Australia (Australia Wool Inovation, 2003). Namun, kain wol konvensional memiliki kecenderungan pil yang relatif tinggi, yang telah berkontribusi terhadap penurunan pangsa wol di pasar serat dunia (Australia Wool Inovation, 2003).

Baru-baru ini, sebuah proses untuk memproduksi kain wol bukan tenunan telah dikomersialkan di Australia. Proses nonwoven 30 persen lebih murah dan 30 kali lebih cepat dari kain wol tradisional dengan menghilangkan tahap pemintalan dan pertenunan konvensional (Australia Wool Inovation, 2003). Masuknya wol menjadi aplikasi nonwoven akan menciptakan pasar baru untuk wol Australia sebagai pesaing untuk kain *nonwoven*. Namun, keberhasilan pakaian bukan tenunan semacam itu akan terbatas sampai titik tertentu, yakni bergantung pada kecenderungan pilling kain *nonwoven* tersebut. Sampai saat ini, hampir tidak ada penelitian yang dipublikasikan mengenai mekanisme, pengukuran, prediksi dan pengendalian pilling pada kain wol *nonwoven*, dan masalah ini akan sangat penting dalam menentukan kebergunaan wol dalam banyak aplikasi tekstil *nonwoven*.

Perkembangan tekstil wol *nonwoven* secara praktik dan komersial merupakan inovasi yang signifikan, yaitu menciptakan "kain dengan sifat unik yang tidak dapat dicapai dengan rajutan atau tenun tradisional, serta membuka berbagai peluang pasar baru untuk wol Merino" (Wool Research Organisation, 2003). Pasar utama untuk produk wol *nonwoven* Australia bersifat internasional, dan ekspor komersial produk ini merupakan strategi utama dalam pengembangan *nonwoven* (Dockery, 2003). Untuk potensi kain wol *nonwoven*, dan pakaian jadi khususnya, untuk merealisasikan masalah *pilling* wol perlu diatasi (The MathWorks Inc, 2000). Australian Wool Innovation (AWI) telah mengidentifikasi bahwa penghilangan *pilling* adalah pesan utama dari konsumen, pengecer dan perancang produk tekstil (Australia Wool Inovation, 2002). Pengurangan pilling juga terdaftar sebagai prioritas utama (Australia Wool Inovation, 2003).

*Fabric pilling* adalah masalah serius bagi industri pakaian jadi (Okpowman, 1998). Pil menyebabkan penampilan tak enak dipandang dan bisa menyebabkan kerusakan dini pada kain (Ramgulam, 1993). Elemen kunci dalam pengendalian *pilliing* kain adalah evaluasi ketahanan terhadap *pilling* dengan pengujian. Ketahan terhadap *pilling* biasanya diuji di laboratorium dengan proses yang mensimulasikan gosokan pada kain, diikuti dengan penilaian manual tingkat *pilling* oleh penguji berdasarkan perbandingan visual sampel dengan satu set gambar uji (Abril, 1998). Keluhan yang sering terjadi tentang metode evaluasi manual atau visual adalah ketidakkonsistenan dan ketidakakuratan hasil *grading* (Xu, 1997). Dalam upaya untuk membawa lebih banyak objektivitas ke dalam proses *grading*, sejumlah sistem otomatis berdasarkan analisis citra telah dikembangkan dan dijelaskan dalam literatur (Abril, 1998, Amirbayat, 1994, Hsi, 19898, Sirikasemleert, 2000 dan Xu, 1997). Semua metode yang ada ini menggunakan peralatan yang mahal dan rumit, seperti proses pencitraan dengan prinsip triangulasi laser (Ramgulam, 1993, dan Sirikasemleert, 2000), dan/atau menggunakan algoritma pengolahan citra kompleks yang melibatkan beberapa tahap (Abril, 1998 dan Xu, 1997).

Palmer (2005) telah mempelopori metode pemodelan penilaian *pilling* objektif yang inovatif dan sederhana berdasarkan pada kuantitatif standar deviasi nilai pixel pada gambar anyaman menggunakan *two dimensional discrete wavelet transformation* (2DDWT). Gambar 1 menunjukkan: a) tiga dari lima gambar pengujian evaluasi *pilling* standar dari rangkaian jaket James H. Heal &

Company Limited '1840, berikut dengan penilaian *supplier pilling*-nya (5 = *unpilled*, 3 = *moderately pilled*, 1 = *heavily pilled*); b) distribusi dari 2DDWT pada empat tingkat analisis menggunakan wavelet Haar; dan c) plot intensitas pemodelan citra pilling terhadap standar deviasi dari koefisien 2DDWT empat tingkat.

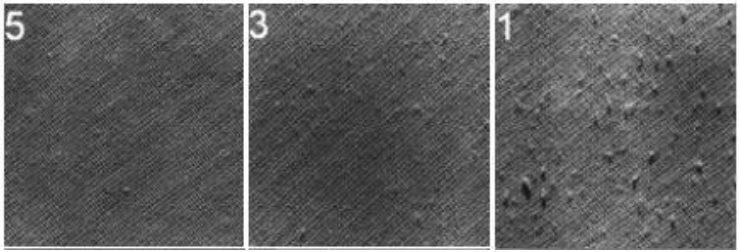

(a)

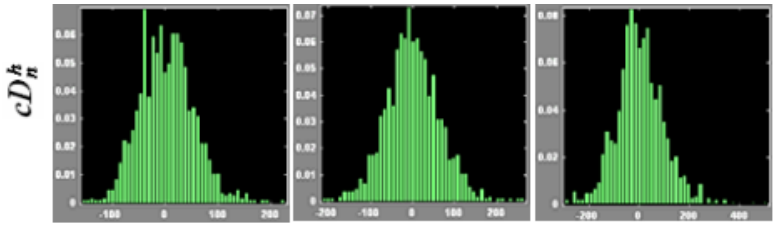

(b)

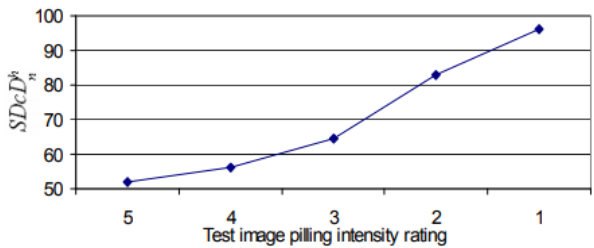

(c)

Gambar-16 Citra permukaan kain (a), hasil distribusi 2DDWT (b) dan grafik standar deviasi dengan analisis wavelet

Palmer (2005) telah mengembangkan metode heuristik untuk memilih parameter analisis wavelet yang optimal (Palmer, 2003) dan menetapkan bahwa metode ini mampu untuk menerjemahkan sampel yang diuji (lihat Gambar-17) dan variasi dalam iluminasi sampel di bawah uji (lihat Gambar-18) (Palmer, 2004). Penerapan analisis wavelet terhadap deteksi otomatis kekurangan kain adalah hal yang baru (Hu, 2000, Latif, 2001, Li, 2002, Sari, 1999 dan Wen, 2001). Teknik yang mendasari analisis wavelet yaitu merupakan pendekatan baru untuk mengatasi permasalahan pada penilaian *pilling* secara obyektif dengan menggunakan analisis citra.

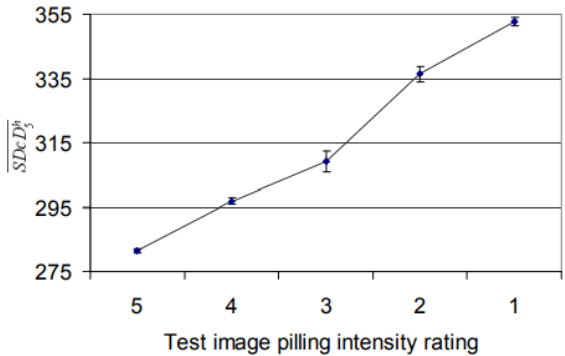

Gambar-17 Rata-rata dari simpangan baku analisis wavelet

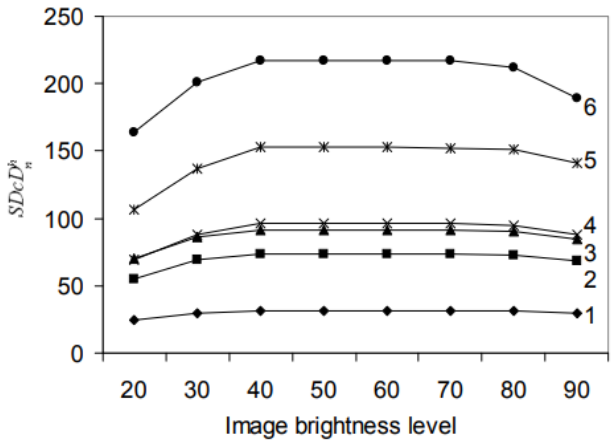

Gambar-18 Simpangan baku pada analisis wavelet

Ketahanan terhadap pilling biasanya diuji dengan memakai suatu uji gosok, dilanjutkan dengan penilaian manual terhadap tingkat *pilling* berdasarkan perbandingan visual sampel dengan satu set gambar uji. Ada satu set gambar tes standar internasional yang didasarkan pada kain wol *nonwoven*, yaitu seperangkat gambar standar 'SM50 Blanket' milik Woolmark. Gambar ini menyediakan empat sampel representatif untuk masing-masing lima tingkat intensitas pilling Gambar-19 menunjukkan salah satu sampel representatif untuk tiga dari lima gambar uji evaluasi pilling standar dari set 'SM50 Blanket' Woolmark, termasuk intensitas dari pilling (5 = tidak ber-*pilling*, 3 = cukup ber-*pilling*, 1 = sangat ber-*pilling*). Seperangkat gambar uji ini akan digunakan sebagai dasar pengembangan teknik analisis citra berbasis wavelet untuk menilai secara obyektif intensitas pilling untuk kain wol *nonwoven*.

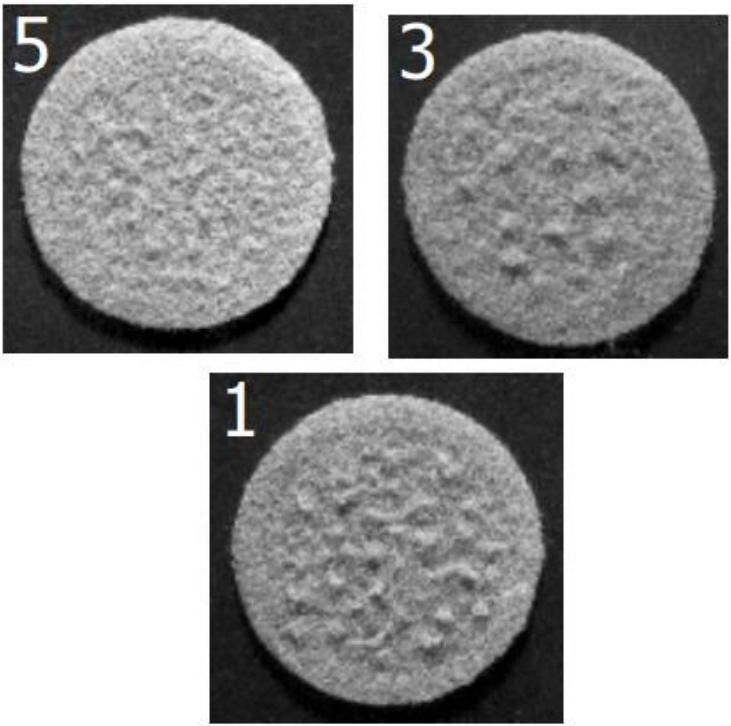

Gambar-19 Kenampakan dari masing-masing skala *pilling* SM50 Blanket

Proses transformasi wavelet diskrit dua dimensi menghasilkan dua komponen analisis, yaitu koefisien detail dan koefisien aproksimasi. Koefisien detail mewakili komponen frekuensi spasial yang tinggi dari citra, dan merupakan dasar yang digunakan oleh peneliti untuk mengkarakterisasi dampak pilling pada struktur periodik (struktur rajutan atau tenun) yang ada pada kain tenun (Palmer, 2003). Untuk kain *nonwoven*, peneliti mengusulkan bahwa struktur aperiodik acak dari kain dapat dicirikan oleh koefisien perkiraan wavelet, yang mewakili komponen frekuensi spasial rendah pada gambar. Peneliti mengusulkan bahwa akan ada skala analisis wavelet yang akan membedakan antara struktur *nonwoven* acak yang mendasari dan adanya struktur pil yang lebih besar pada sampel kain, dan bahwa distribusi koefisien perkiraan wavelet pada skala analisis tersebut akan memberikan ukuran kuantitatif pada *pilling* kain.

Perangkat Woolmark 'SM50 Blanket' merupakan gambar pilling standar menyajikan empat contoh masing-masing dari lima tingkat intensitas pilling. 20 gambar ini dipindai pada 600 titik per inci dan dipotong. Sedangkan penelitian Palmer (2003), telah terbukti kuat terhadap variasi kecerahan gambar mempengaruhi hasil pengujian. Ada banyak aplikasi pengolahan citra yang sensitif terhadap variasi kecerahan gambar (Ghassemieh, 2002). Dalam contoh ini, peneliti mengusulkan untuk menggunakan koefisien perkiraan wavelet sebagai dasar analisis, karena koefisien aproksimasi mewakili informasi frekuensi rendah pada gambar, sehingga hasilnya akan sensitif terhadap variasi kecerahan gambar (Mandal, 1999). Nilai pemerataan histogram *pixel* gambar adalah metode yang berguna untuk menempatkan gambar dalam format yang konsisten sebelum dilakukan proses perbandingan (Castleman, 1996), dan dilaporkan dalam aplikasi analisis gambar wavelet (Mojsilovic, 1997) dan lainnya (Srisuk, 2001) sebagai teknik untuk menangani dengan variasi kecerahan gambar. Pada 20 gambar standar uji tadi, telah dilakukan proses pemerataan nilai histogram *pixel* untuk menangani variasi kecerahan warna pada gambar.

Untuk masing-masing dari 20 gambar standar ini, empat gambar tambahan disintesis dengan memotong satu tepi gambar standar sekitar 15 persen. Ini menghasilkan 100 gambar secara total; 20 untuk setiap intensitas *pilling*. Untuk masing-masing dari 100 gambar, standar deviasi dari distribusi koefisien perkiraan ($SDcA_n$) pada berbagai skala analisis. Berdasarkan analisis dengan menggunakan wavelet Haar, $SDcA_n$ dihitung dengan menggunakan Matlab *Wavelet Toolbox* (The MathWorks Inc, 2004). Dengan menggunakan nilai mean $SDcA_n$ yang diperoleh untuk 20 citra uji pada setiap tingkat intensitas pilling. Telah ditemukan bahwa analisis 2DDWT pada lima skala menghasilkan hubungan monoton antara intensitas *pilling* dan $SDcA_5$. Gambar-20 menyajikan nilai rata-rata dan interval kepercayaan 90 persen untuk $SDcA_5$ (standar deviasi dari distribusi koefisien perkiraan wavelet untuk analisis tingkat 5) untuk masing-masing intensitas *pilling*.

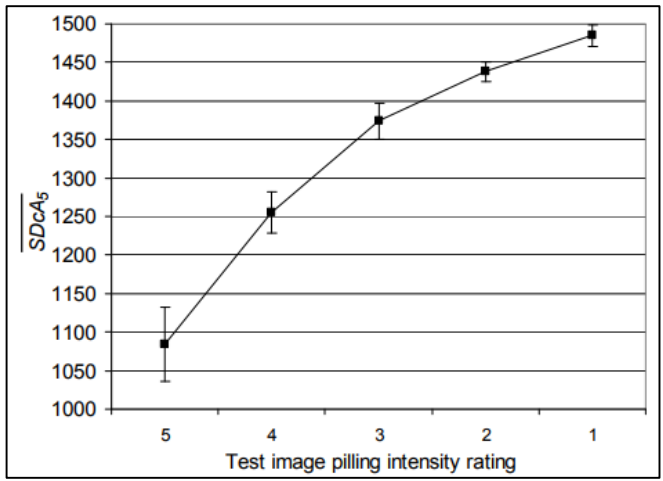

Gambar-20 Rata-rata simpangan baku dari nilai koefisien aproksimasi wavelet level 5 dan interval kepercayaan 90%-nya untuk gambar standar uji SM50 Blanket

Berdasarkan hasil analisis tersebut, dimungkinkan untuk menerapkan metode analisis citra ini ke sekumpulan sampel standar kain untuk mengembangkan kurva karakteristik yang dikalibrasi yang menghubungkan intensitas pilling dengan $SDcA_n$ yang diperoleh dengan analisis sampel uji kain. Dengan cara ini adalah mungkin untuk melakukan evaluasi terhadap intensitas *pilling* yang serupa dengan metode perbandingan visual, namun setelah dikalibrasi untuk jenis kain dan lingkungan uji tertentu, akan menghasilkan ukuran yang obyektif tanpa interpretasi manusia. Berdasarkan penggunaan gambar uji *pilling* standar 'SM50 Blanket' milik Woolmark untuk kain *nonwoven* dan konsep analog dari karya penulis sebelumnya, kelayakan teknik analisis baru untuk menghubungkan standar deviasi distribusi koefisien aproksimasi 2DDWT dengan intensitas pilling kain *nonwoven* terbukti.

## 5. PENDETEKSIAN DAN PENGKLASIFIKASIAN CITRA WEB NONWOVEN MENGGUNAKAN NEURAL NETWORK

Inspeksi visual merupakan bagian penting dalam pengendalian kualitas di industri. Kontrol kualitas dirancang untuk memastikan bahwa produk yang cacat tidak lolos dan berada ditangan pelanggan. Untuk alasan ini, pemantauan karakteristik produk dan penampilan selama produksi sangat penting. Sampai beberapa tahun terakhir, pekerjaan ini sangat bergantung pada inspeksi manusia. Tapi hari ini semua pabrik manufaktur menuju sistem kontrol kualitas otomatis. Fakta ini masuk dalam industri tekstil dan *nonwoven*. *Nonwoven* adalah salah satu cabang baru di industri tekstil yang dalam 25 tahun terakhir, jumlah tekstil *nonwoven* yang digunakan untuk aplikasi industri dan komersial meningkat lebih dari 10 kali. Berbagai macam produk industri, mulai dari karpet hingga serbet bayi memanfaatkan kain *nonwoven*. Untuk semua produk ini, kualitas kain memegang peranan penting. Pentingnya penampilan *nonwoven* dan juga pengaruh yang ketidakkeseragaman *web nonwoven*, dan kerugian ekonomi yang mengikutinya, sehingga pengawasan dan pemantauan kualitas secara langsung menjadi hal yang sangat penting. Penerapan algoritma pengolahan citra ke dalam pengembangan peralatan cepat dan khusus telah menyelesaikan masalah nyata pada dunia inspeksi industri. Sistem *industry vision* harus beroperasi secara *real-time*, menghasilkan sedikit jumlah tingkat alarm palsu dan fleksibel untuk mengakomodasi variasi di lokasi inspeksi.

Pemeriksaan visual otomatis bahan web adalah tugas yang sangat kompleks dan penelitian di bidang ini masih sangat lebar untuk diteliti (Brzakovic & Vujovic 1996). Banyak usaha telah dilakukan untuk memecahkan masalah ini. Norton-Wayne dkk (1992) telah mendeskripsikan suatu sistem sederhana , berdasarkan *adaptive thresholding* dan *biner filtering*. Sistem optik untuk mendeteksi cacat secara *real-time* ditunjukan pada penelitian Olsson dan Gruber (1993). Hal ini didasarkan pada hamburan cahaya dan menggunakan peralatan elektro-optik untuk mendeteksi cacat. Dar dkk (1997) menyajikan sebuah sistem untuk mendeteksi satu jenis cacat (*pilling*) di lima tingkat. Transformasi Radon digunakan untuk ekstraksi fitur dan logika fuzzy yang diimplementasikan untuk penilaian. Escofet dkk. (1996) menganalisis berbagai cacat pada kain yang berbeda dan dalam setiap kasus, sementara fungsi Gabor digunakan untuk ekstraksi fitur. Mueller dan Nickolay (1994) menggunakan pengolahan citra morfologi untuk inspeksi tingkat abu-abu. Sistem Huart dan Postaire (1994) telah menggunakan multi-kamera dengan perangkat keras terkait. Kelompok penelitian di institut teknologi Georgia (1997) menerapkan transformasi wavelet dan logika fuzzy untuk menyelesaikan tugas ini. Stojanovic dkk (1999) membuat suatu sistem sederhana, berdasarkan algoritma biner cepat untuk menentukan daerah cacat yang mungkin dan menggunakan *artificial neural network* untuk mengklasifikasikan cacat. Campbell pada tahun 1997 menghadirkan Transformer Fourier Diskrit 2D untuk menampilkan ekstraksi pada *artificial neural network* pada proses klasifikasi. Pakkanen & et al. (2004) menggunakan deskripsi fitur standar MPEG-7 dan fitur *tree organizing* untuk mengklasifikasi citra dari bagian cacat. Ahmet (2000) menggunakan matriks *cooccurrence* untuk ekstraksi dan pemeriksaan cacat. Serdaroglu dkk (2004) menggunakan wav*elet package transform* untuk mengekstraksi fitur gambar dan jarak Euclidean untuk mendeteksi cacat. Karras dkk (1998) menyajikan metode ekstraksi fitur yang didasarkan pada transformasi *wavelet*, analisis SVD, dan matriks *co-occurrence*. Sezer dkk (2003) telah menerapkan analisis komponen independen (ICA) untuk ekstraksi fitur dari suatu citra. Pada penelitian yang dilakukan oleh Yousefzadeh dkk (2006), dimensi fractal dari citra *nonwoven*, digunakan dalam hal ekstraksi fitur citra. *Artificial neural network* beserta *back-propagation algorithm* telah diimplementasikan untuk mendeteksi dan mengklasifikasikan cacat pada kain *nonwoven*. Dengan menggunakan beberapa jumlah cacat yang mungkin terdeteksi, sistem pengukuran telah berhasil memberikan hasil yang konsisten pada eksperimen dan pada implementasi di industri.

Pada eksperimen yang dilakukan oleh Yousefzadeh dkk (2006), sampel *nonwoven* dari proses sistem *thermalbonding* telah digunakan untuk proses investigasi dan analisis. Spesifikasi sampel yang digunakan dapat dilihat pada Tabel-2.

Tabel-2 Spesifikasi bahan *nonwoven*

| No | Sifat fisik dan mekanik | Nilai |
|---|---|---|
| 1 | Material | 100% polipropilen |
| 2 | Kerapatan | $18g/m^2$ |
| 3 | Ketebalan | 120 mikron |
| 4 | Mulur ke arah sejajar dengan arah mesin | 28% |
| 5 | Kekuatan basah ke arah sejajar dengan arah mesin | 29N |
| 6 | Kekuatan kering ke arah sejajar dengan arah mesin | 30N |
| 7 | Kekuatan basah ke arah tegak lurus dengan arah mesin | 5N |
| 8 | Kekuatan kering ke arah tegak lurus dengan arah mesin | 6,2N |

Pemindai yang digunakan untuk menangkap citra digital *nonwoven* diatur pada resolusi terbaik. Resolusi pemindaian diatur pada 200 dpi, serta dihindari terbentuknya bayangan pada gambar. Tiga puluh citra telah dipindai pada bagian cacat dan bagian non cacat di kain *nonwoven.* Setelah itu dilakukan proses *thresholding* pada citra, dan citra *grey scale* diubah kembali ke dalam gambar hitam putih (citra biner). Nilai *thresholding* tersebut diperoleh dari histogram citra digital *nonwoven* dan dari nilai rata-rata serta varians nilai skala keabuan pada citra *nonwoven*. Pada Gambar-21 menunjukan contoh dari bagian cacat dan bagian non-cacat pada kain *nonwoven*.

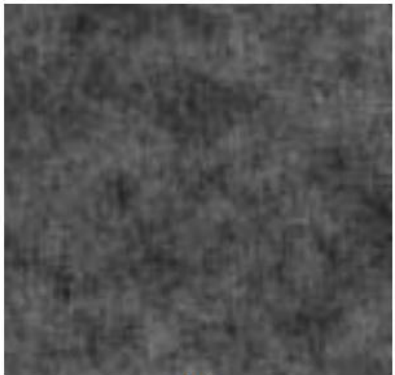
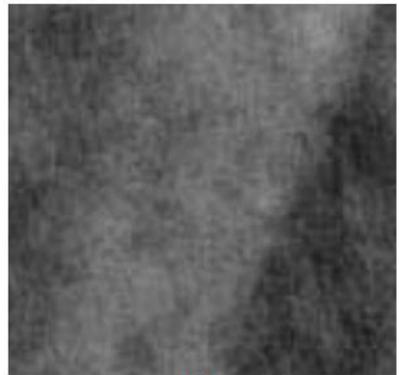
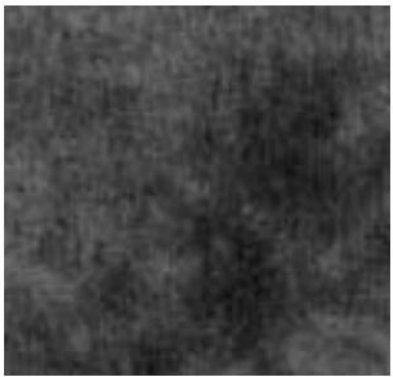
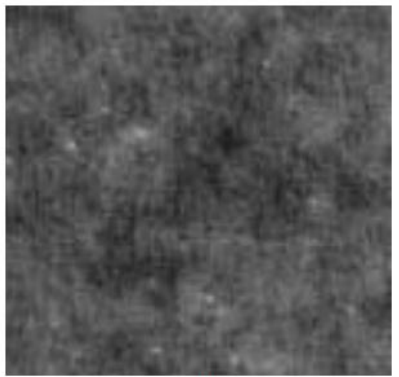

Gambar-21 Contoh dari citra kain *nonwoven* sebelum dilakukan proses pengolahan citra, bagian non-cacat (a), bagian *thick spot* (b), bagian *thin spot* (c) dan bagian *neps* (d)

Citra digital dari kain nonwoven diseragamkan ukurannya dalam 128 x 128 pixel, serta nilai rata-rata dari nilai keabuan citra dihitung. Proses *thresholding* dilakukan pada citra digital dan menetapkan ambang batas nilai citra pada gambar (nilai tersebut diperoleh pada bagian histogram pada citra *grey scale*, kemudian citra diubah ke dalam bentuk citra biner. Gambar-22 menunjukan hasil citra biner yang berasal dari citra asli pada Gambar-21.

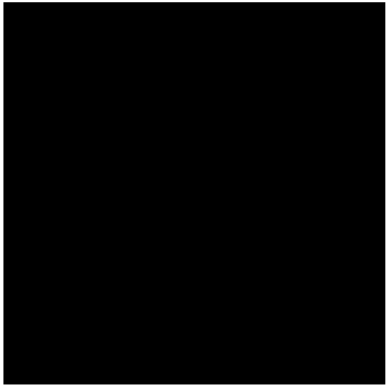
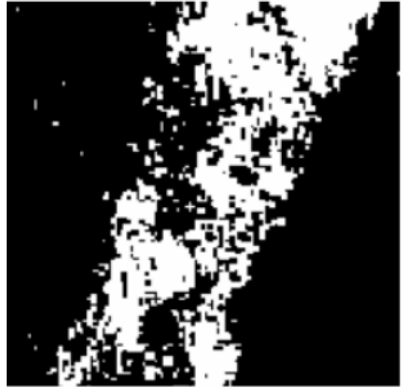
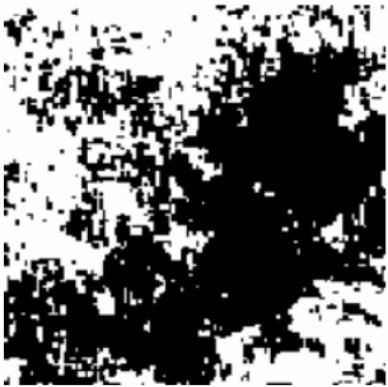
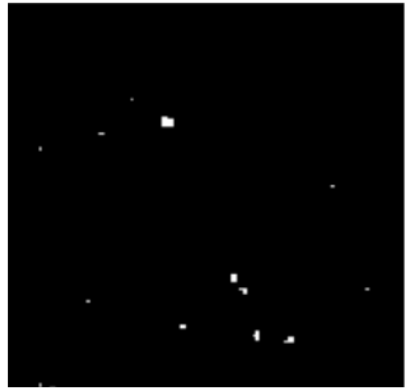

Gambar-22 Citra digital setelah pengolahan citra digital, bagian non-cacat (a), bagian *thick spot* (b), bagian *thin spot* (c) dan bagian *neps* (d)

Setelah mengubah citra menjadi hitam dan putih, *box counting* pada dimensi fractal dapat didefinisikan. Grafik pada Gambar-23 menunjukan hasil dari penerapan algoritma fractal pada salah satu citra cacat.

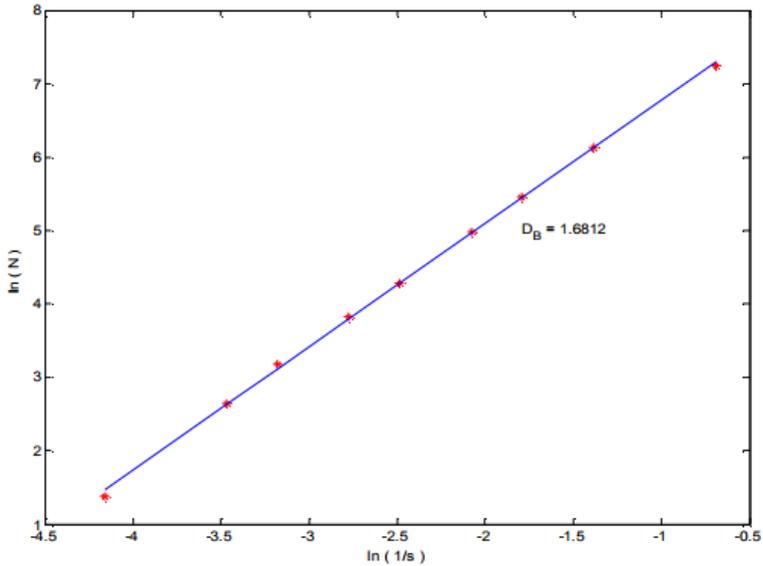

Gambar-23 Grafik hasil prosedur *box counting* pada citra

Pada penggunaan dimensi fractal, kerapatan gambar juga dapat mempengaruhi. Dimensi fraktal didefinisikan sebagai persentase dari *pixel* putih dan *pixel* hitam. Pada Tabel-3 menunjukan nilai hasil dari sampel yang berasal dari ilustrasi Gambar-21.

Tabel-3 Hasil penerapan pengolahan citra dan dimensi fraktal

| Sampel | Rata-rata (pixel) | Varians | Std | Kerapatan | Fraktal |
|---|---|---|---|---|---|
| 1 | 81.04 | 48.28 | 6.95 | 0 | |
| 2 | 89.45 | 92.92 | 9.64 | 25.18 | 1.6812 |
| 3 | 67.16 | 75.51 | 8.69 | 42.11 | 1.8266 |
| 4 | 84.65 | 60.84 | 7.79 | 0.23 | 0.4101 |

Setelah proses ekstraksi dan penentuan berbagai nilai dari citra (seperti dimensi fractal, kerapatan dan citra *gray scale*, telah dibuat dua buah lapis *neural network* yang didesain untuk mengolah parameter-parameter dari citra sebagai sebuah input. Jumlah lapisan dan *neuron* ditentukan dengan metoda *trial and error*. Struktur *network* dapat dilihat pada Gambar-24. Empat pengklasifikasian telah dibuat, yaitu:

1. Non-cacat     (a => kode: 1000)
2. *Thick spot*  (b => kode: 0100)
3. *Thin spot*   (c => kode: 0010)
4. *Neps*        (d => kode: 0001)

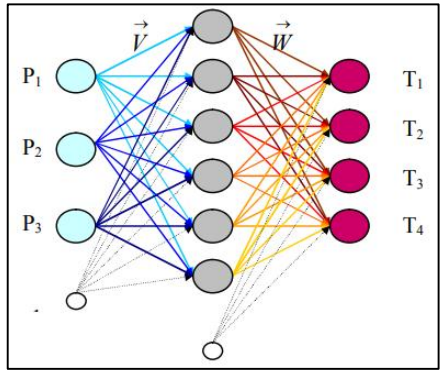

Gambar-24 Struktur dari *neural network*

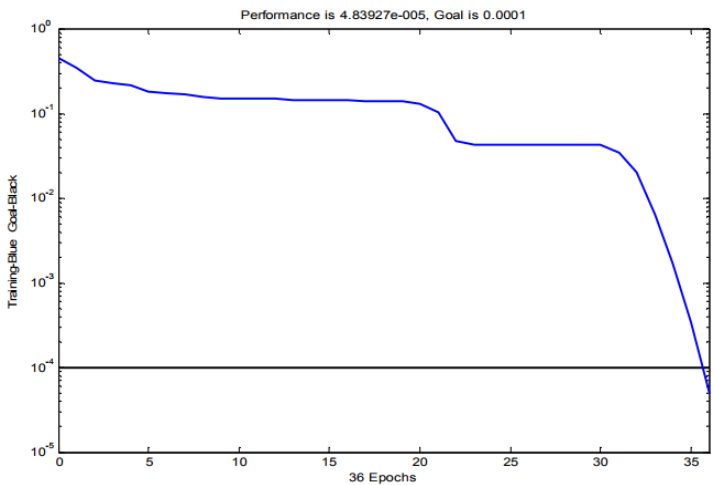

Gambar-25 Hasil MSE pada salah satu citra *nonwoven*

# 6. ANALISIS CITRA DIGITAL UNTUK MENENTUKAN DISTRIBUSI UKURAN PORI-PORI DARI KAIN GEOTEXTILE NONWOVEN

*Geotextiles* adalah bahan permeabel yang terdiri dari plastik polimer. Geotekstil digunakan dalam berbagai aplikasi geoteknik. Salah satu fungsi utamanya adalah fungsi filtrasi yang memungkinkan aliran air dapat mengalir namun menahan migrasi tanah melintasi bidang *geotextile* tersebut. Ada dua kelompok geotekstil utama berdasarkan metode pembuatannya, yaitu dengan menggunakan tenunan (anyaman tekstil) dan serat *nonwoven*. Sejumlah metode telah dikembangkan untuk menentukan distribusi ukuran pori dari geotekstil *nonwoven*, hal tersebut yang penting untuk menentukan fungsinya sebagai media saring, yaitu potensi penyumbatan dan retensi tanah saat kontak dengan berbagai tanah. Namun sebagian besar metode saat ini hanya memberikan cara tidak langsung untuk menentukan PSD (*Pore size distribution*) dan masih terdapat kelemahan. Terdapat dua metode yang dapat digunakan untuk menentukan PSD, yaitu (1)uji pengayakan kering dan (2) analisis citra, uji pengayak kering lebih umum digunakan yang tercantum pada standar ASTM D 4751. Uji pengayak kering memberikan cara langsung untuk mengukur ukuran bukaan yang jelas (AOS, *Apparent opening size*) (Giroud 1996). Namun, penelitian terbaru menunjukkan bahwa ukuran pembukaan pori yang lebih kecil dapat mempengaruhi kinerja filtrasi geotekstil secara signifikan, dan kurva PSD yang lengkap harus ditentukan (Fischer dkk, 1990, Fischer 1994, Aydilek 2000). Uji pengayak kering jauh dari menyediakan kurva yang lengkap karena keakuratan uji untuk ukuran pembukaan pori yang lebih kecil dari 90 $\mu$m.

Metode analisis citra jarang digunakan dalam penetapan PSD geotextiles *nonwoven* karena kurangnya prosedur yang telah terbukti. Sebagian besar metode analisis citra yang ada didasarkan pada pendekatan dua dimensi dengan menggunakan pandangan planar geotekstil (di bidang geotekstil) dan secara manual menghitung jumlah ukuran pembuka pori yang berbeda. Dengan demikian, kurva PSD yang diperoleh dengan pendekatan ini adalah PSD dari permukaan planar yang dianalisis yang tidak mewakili variasi ukuran pembukaan pori melalui ketebalan geotekstil. Adapun beberapa metode menggunakan pandangan penampang melintang geotekstil, yaitu normal terhadap bidang geotekstil (Masounave dkk, 1980, Elsharief dan Lovell 1996). Namun, metode analisis citra ini sebagian besar bersifat operator dependen.

Penelitian yang ada telah menunjukkan bahwa ada kebutuhan untuk mengembangkan pendekatan yang lebih baik dalam menentukan PSD geotekstil (Rollin dkk, 1982, Lombard dan Rollin 1987, Smith 1993, Bhatia dkk, 1993, Elsharief dan Lovell 1996, Dierickx, 1999). Sebagai tanggapan atas kebutuhan ini, metode penentuan PSD berbasis gambar baru telah dikembangkan untuk geotekstil nonwoven dan disajikan pada penelitian Ahmet (2002). Metode ini menggunakan pandangan planar dan penampang melintang untuk menangkap struktur 3D dari geotekstil nonwoven. Metode terdiri dari tiga tahap, yaitu persiapan spesimen, analisis citra, dan penentuan ukuran bukaan pori. Metode analisis citra dikembangkan dengan menggunakan berbagai algoritma morfologi matematis untuk mengetahui ukuran pembukaan pori geotextile. Terdapat dua karakteristik ukuran pembukaan pori-pori, yaitu $O_{95}$ dan $O_{50}$, yang ditentukan dengan menggunakan informasi yang diberikan oleh analisis citra. Akhirnya, nilai yang terukur diperiksa terhadap nilai AOS yang dilaporkan oleh pembuat dan nilai dari dua karakteristik ukuran pembukaan pori berdasarkan uji laboratorium dan persamaan analitis.

Dua belas jenis geotekstil *nonwoven* telah dianalisis, namun hanya lima yang mencakup AOS $O_{95}$. Geotekstil dipilih dari yang paling sering digunakan dalam aplikasi filter. Dua dari lima sebelumnya telah digunakan oleh Smith (1993) untuk mengevaluasi berbagai metode penentuan PSD. Sifat fisik dan hidrolik geotextiles ditunjukan pada pada Tabel-4.

Struktur 3D dari geotextiles nonwoven menimbulkan kesulitan dalam menangkap struktur pori dari gambar 2D. Bagian tipis planar dan penampang melintang diperlukan untuk memberikan informasi rinci pada kain geotekstil. Bagian tipis dari lima geotextiles yang diteliti dalam penelitian ini disiapkan dan mengikuti prosedur yang umumnya digunakan untuk pembuatan lapisan tipis pada tanah dan batuan.

Tabel-4 Sifat dari bahan geotekstil *nonwoven* yang digunakan pada penentuan PSD

| Geo-tekstil | Struktur, jenis polimer | Gra-masi (g/m$^2$) | Kete-balan (mm) | *Apparent opening size* (mm) | Poro-sity (%) | Permiti-vitas (s$^{-1}$) |
|---|---|---|---|---|---|---|
| N | Nonwoven, needle-punched, heat-bonded, polypropilene | 136 | 0.45 | 0.28 | 66.4 | 0.70 |
| P | Nonwoven, needle-punched, staple-fiber, polypropilene | 387 | 3.0 | 0.106 | 85.7 | 0.80 |
| M | Nonwoven, needle-punched, continous filament, polypropilene | 340 | 2.53 | 0.15 | 85.0 | 1.10 |
| C4 | Nonwoven, needle-punched, continous filament, polypropilene | 401 | 2.92 | 0.15 | 84.7 | 1.0 |
| D1 | Nonwoven, needle-punched, staple-fiber, polypropilene/polyester | 228 | 2.21 | 0.075-0.104 | 88.5 | 1.35 |

Catatan: nilai porositi telah ditentukan dengan menggunakan metoda yang dideskripsikan oleh Wayne dan Koerner (1994). Nilai terendah dari AOS yang dilaporkan didasarkan pada laporan pembuat kain. Seluruh sifat merupakan detail laporan pembuat kain, kecuali untuk geotekstil C4 dan D1 yang ditentukan oleh Smith (1993).

Dari masing-masing tipe geotextile, tiga spesimen disiapkan, dengan masing-masing spesimen termasuk tiga penampang melintang dan satu sampai dua bagian penampang planar (bagian planar merupakan bagian yang berada pada dasar material geotekstil) (dapat dilihat pada Gambar.-26). Persiapan bagian tipis memerlukan serangkaian langkah berurutan, yaitu impregnasi epoksi-resin, pemotongan, penggerindaan, *lapping*, dan pemolesan (Aydilek 2000). Resin viskositas rendah dan pengeras resin digunakan untuk impregnasi geotekstil bukan tenunan. 25 x 25 mm contoh uji geotextile dikeringkan pada suhu kamar 20°C selama 24 jam dan ditempatkan dalam cetakan persegi dengan permukaan bawah rata. Lima bagian berat resin dicampur dengan satu bagian berat pengeras, dan campuran perlahan dipindahkan ke dalam cetakan dengan menggunakan semprit. Waktu pengeringan campuran resin epoksi memakan waktu kira-kira selama 7-8 jam pada suhu kamar.

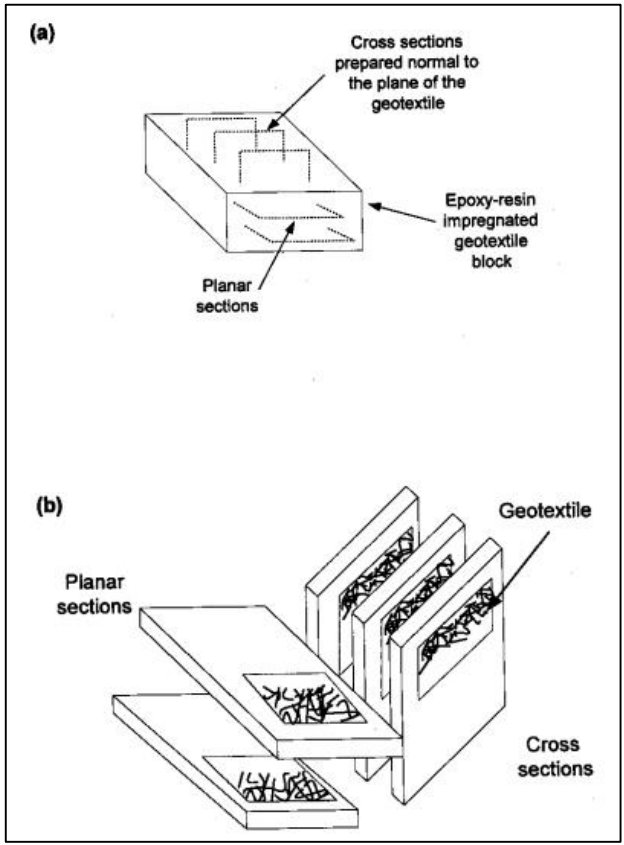

Gambar-26 Persiapan contoh uji, penempatan bahan geotekstil pada bahan cor resin epoksi (a) dan proses pemotongan resin epoxy

Blok kering diambil dari cetakan dan dipotong oleh pemotong pisau berlian Buehler. Blok yang diimpregnasi kemudian digrinda sampai kira-kira 25 mm menggunakan serangkaian cakram penggerinda berlian 125, 70, dan 30 mm. Blok kemudian dilekatkan pada slide mikroskop menggunakan epoxy. Sebuah pemotong Lapar-Ingram digunakan untuk memotong balok-balok di sepanjang bidang yang ditunjukkan pada Gambar-26. Air digunakan sebagai cairan pemotong. Ketebalan akhir bagian adalah sekitar 700 mm setelah dua seri pemotongan. Bagian yang dipotong digerinda sampai ketebalan 450 mm menggunakan penggerinda bagian Wards-Ingram. Proses *lapping*, kadang-kadang disebut sebagai ''*planar grinding*'', digunakan untuk menghilangkan bekas yang tertinggal dari permukaan hasil proses *grinding*. Bubuk gril silikon karbida, berukuran 17,5 mm, digunakan sebagai abrasif untuk proses *lapping* di mesin *lapping* produksi Logitech LP-30.

Gambar struktur pori dari geotextiles *nonwoven* ditangkap menggunakan mikroskop cahaya optik. Mikroskop memiliki *platform workstation* 30 x 25 mm dan lensa zoom makro 0,7x sampai 7X yang digabungkan dengan program pengambilan gambar yang disebut Pixeria. Mikroskop memberikan tingkat pembesaran hingga 140X. Kamera digital Olympus (C-35AD-2) Dilekatkan pada mikroskop yang mengirim gambar yang diambil ke Pixeria. Rasio zoom 5 dan 2,75 dan resolusi 1.260 x 960 piksel dan 640 x 480 piksel digunakan untuk gambar planar dan cross-sectional. Ukuran piksel yang sesuai adalah sekitar 4.831023 mm dan 9.431023 mm untuk gambar planar dan penampang melintang. Pembesaran rasio 0.5X untuk semua spesimen. Spesimen disinari dari bawah. Intensitas cahaya disesuaikan sehingga piksel latar belakang memiliki nilai abu-abu 255, yaitu putih murni. Gambar-27 menunjukkan gambar planar dan Gambar-28 penampang lintang geotekstil P yang ditangkap dengan metode ini.

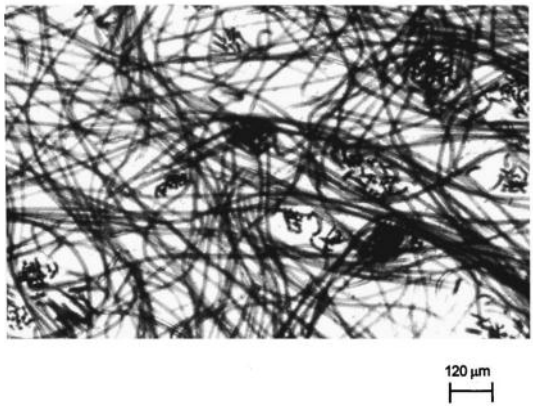

Gambar-27 Citra geotekstil pada bidang planar

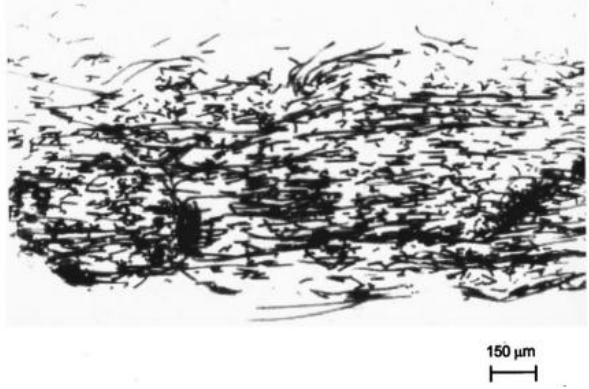

Gambar-28 Citra geotekstil pada bidang penampang melintang

Dalam pengolahan citra, *thresholding* sangat penting dan berguna untuk membedakan latar belakang dari objek gambar yang diminati. Nilai histogram pixel dari gambar geotekstil nonwoven menunjukkan perilaku bimodal dari setiap elemen pada gambar (dapat dilihat pada Gambar-29a). Distribusi nilai pixel bimodal ini menunjukkan sifat biner yang melekat pada gambar-gambar ini. Oleh karena itu, keputusan untuk mengembangkan algoritma pemrosesan gambar yang relevan pada gambar geotextile biner konsisten dengan sifat asli gambar harus dilakukan. Selanjutnya, teknik pencitraan skala abu-abu tidak digunakan karena mencoba melakukan pengukuran untuk penentuan PSD pada gambar graylevel dan bukan gambar biner akan meningkatkan kompleksitas komputasi dari algoritma yang diusulkan tanpa kelebihan yang jelas.

Sebuah histogram bimodal menyiratkan bahwa nilai piksel tingkat abu-abu timbul dari campuran dua subpopulasi, dan dengan memilih nilai ambang yang tepat di antara dua puncak distribusi marjinal subpopulasi ini, piksel milik subpopulasi tersebut dapat dibedakan. Subpopulasi terpusat di sekitar nilai piksel yang lebih kecil (contoh, tingkat abu-abu yang lebih gelap) mewakili serat geotextile; sedangkan, subpopulasi terpusat di sekitar nilai piksel yang lebih besar (contohnya, tingkat abu-abu yang lebih terang) mewakili lubang pori. Chow dan Kaneko (1971) telah melakukan pendekatan histogram citra tersebut dengan jumlah dua distribusi Gaussian dan memperoleh ambang batas dengan meminimalkan kesalahan klasifikasi berkenaan dengan nilai ambang yang dipilih. Dalam makalah mereka berikutnya, Chow dan Kaneko (1972) berhasil menerapkan metode ini untuk mendeteksi daerah jantung pada rontgen dada. Nakawaga dan Rosenfeld (1979) menggunakan pendekatan ini untuk menguji citra televisi dari bagian-bagian mesin dan berhasil. Metode ini juga direkomendasikan oleh peneliti lain (Sahoo dkk, 1988, Gonzalez dan Woods, 1992, Henstock dan Chelberg, 1996).

Suatu nilai ambang batas telah digunakan untuk gambar geotekstil nonwoven, nilai tersebut diperoleh dengan menggunakan metode yang dijelaskan oleh Chow dan Kaneko (1971). Dua distribusi Gaussian dipasang pada kurva bimodal yang diamati. Yang paling sesuai dengan kurva diyakinkan dengan meminimalkan kesalahan rata-rata kuadrat data. Titik persimpangan di bagian bawah lembah dipilih sebagai ambang optimal, sesuai saran Kapur dkk, (1985). Gambar-29b menunjukkan penentuan ambang batas dengan metode ini. Sebuah kode pendek ditulis di Matlab untuk mengotomatisasi metode ini.

Histogram beberapa gambar memamerkan nilai yang beragam (Gambar-30a), hal tersebut merupakan masalah utama untuk algoritma *thresholding*. Masalah tersebut disebabkan oleh kombinasi elemen dalam spesimen dan mikroskop yang digunakan. Misalnya, beberapa spesimen geotextile terkontaminasi selama proses penggerindaan. Jika contoh uji tersebut diamati, filter median 1D yang diterapkan pada data histogram digunakan untuk mendapatkan histogram. Filter ini sering digunakan dalam aplikasi pemrosesan sinyal untuk menghilangkan *noise* palsu berupa lonjakan positif atau negatif yang dilapiskan pada sinyal yang diinginkan. Gambar-31a dan Gambar-31b menunjukan gambar geotextile P sebelum dan sesudah operasi *thresholding*.

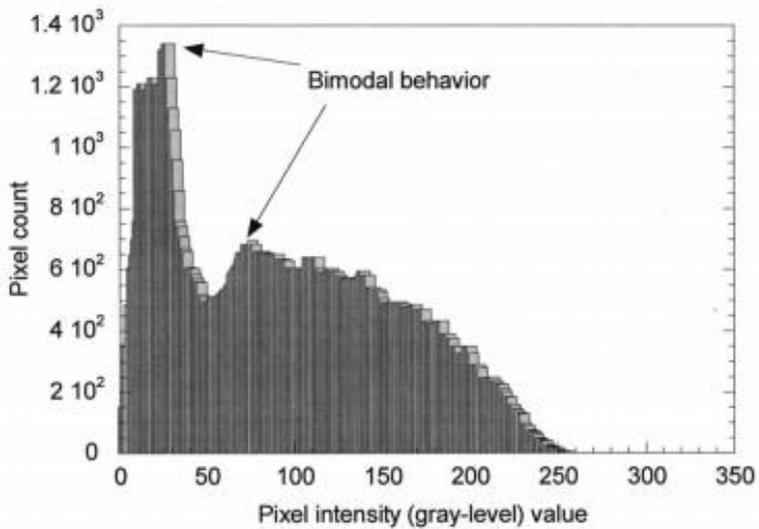

(a)

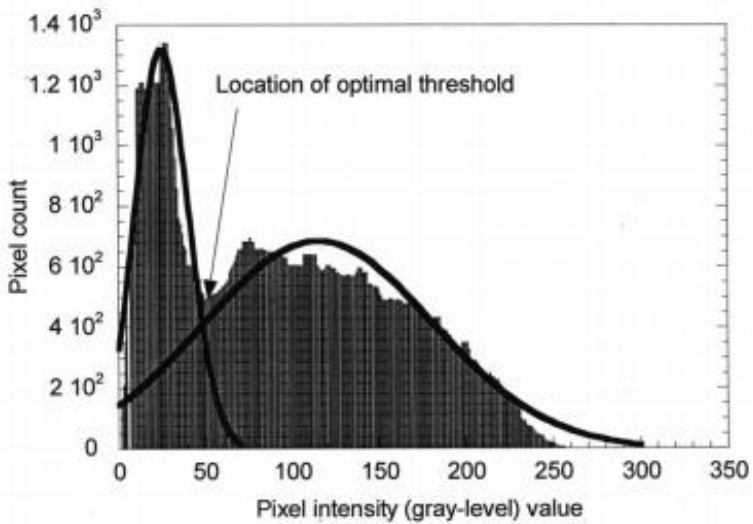

(b)

Gambar-29 Histogram dari citra material geotekstil, (a) histogram dari gambar asli, (b) histogram dari hasil Gaussian Filter

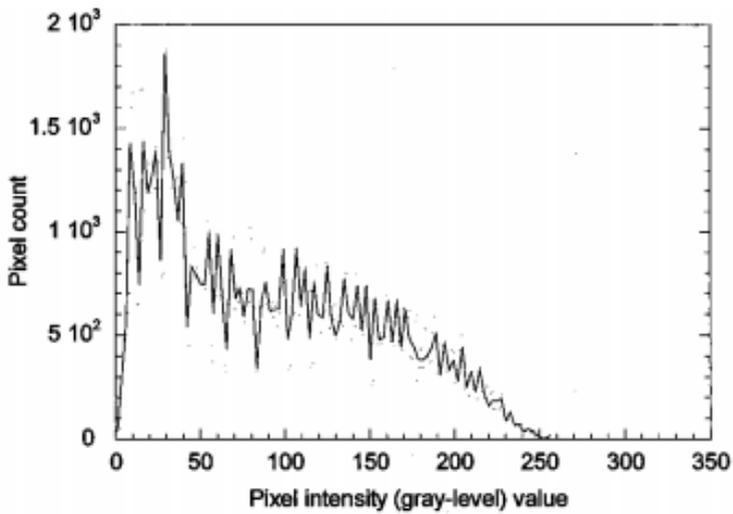

(a)

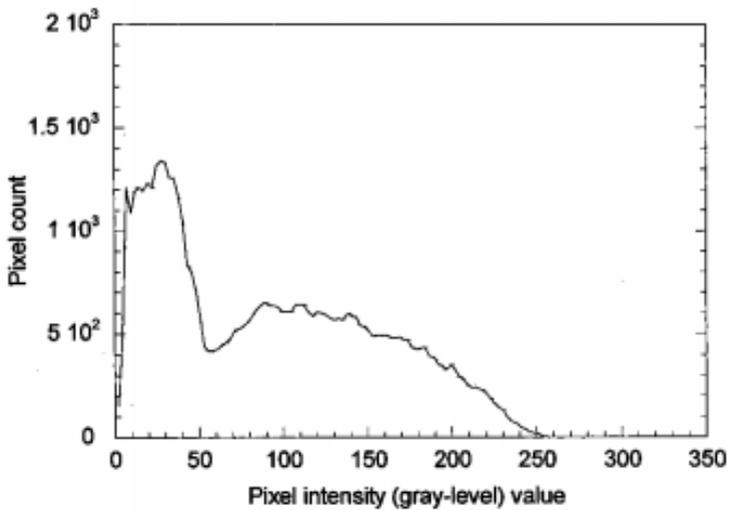

(b)

Gambar-30 Pengaruh *noise* pada gambar, (a) nilai yang bervariasi akibat *noise* dan (b) nilai yang didapatkan setelah dilakukan proses ID *median filter*

Ukuran *noise* yang mewakili objek latar depan dan latar belakang pada gambar *geotextile thresholding* secara fundamental bergantung pada resolusi dimana gambar-gambar ini ditangkap. Jelas, menangkap gambar beresolusi tinggi kemungkinan akan menurunkan tingkat kebisingan. Di sisi lain, pilihan resolusi untuk pengambilan gambar adalah hasil antara perlunya menggambarkan struktur yang akurat (pori-pori dan garis serat geotextile) dan keinginan untuk mengurangi kompleksitas komputasi dan waktu. Nilai resolusi (berbeda untuk potongan potongan planar dan cross-sectional) yang diambil disini dipilih setelah mempertimbangkan hal tersebut. Pada resolusi pengambilan gambar yang diambil, melalui pemeriksaan cermat awal terhadap gambar geotekstil *threshold*, ukuran batas maksimum *noise* yang mewakili objek gambar ditentukan. Ukuran ini masing-masing sekitar 232 dan 333 piksel untuk masing-masing gambar planar dan penampang melintang. Untuk menghilangkan *noise*, elemen penataan isotropik diperlukan. Pemilihan elemen penataan kuadrat merupakan perkiraan diskrit terbaik untuk struktur pada 232 dan 333 grid, yaitu isotropik pada grid sampling 2D yang seragam. Karena noise yang mewakili objek gambar biner (*foreground* dan *background*) tidak menunjukkan bentuk atau orientasi tertentu, elemen penataan isotropik dengan ukuran yang ditentukan berdasarkan ukuran batas maksimum untuk *noise* yang mewakili benda merupakan pilihan terbaik untuk tujuan tersebut. Setiap elemen struktur yang lebih kecil akan menyebabkan meningkatnya tingkat kebisingan residual. Efek distorsi yang diterapkan pada elemen penataan yang tidak tepat sangat sulit diukur secara otomatis karena memerlukan suatu *smart process* untuk membedakan antara fitur *noise* dan *noise-like* (serupa *noise*. Meskipun pemeriksaan dan analisis manual yang membosankan untuk menilai konsekuensi penggunaan elemen penataan lainnya dimungkinkan.

Operasi pembukaan yang digunakan sebagai bagian dari penyaringan morfologi adalah aplikasi sekuensial operasi *erotion* yang diikuti oleh operasi *dilatation* (yaitu proses *opening*, *erosion*, lalu *dilation*), keduanya menggunakan elemen penataan yang sama. *Erotion* menghilangkan *pixel* yang tidak relevan pada gambar dan mengikis kontur objek sehubungan dengan template yang ditentukan oleh elemen penataan. Di sisi lain, *dilatation* (disebut dilatasi) memiliki efek sebaliknya dari erosi, karena melebarkan benda sama dengan mengikis latar belakang. Operasi *opening* menghilangkan semua objek latar depan, dalam hal ini formasi seperti serat, yang sangat kecil dibandingkan dengan SE yang digunakan. Sebagai hasil dari operasi pembukaan, fitur noise yang tampak seperti potongan serat kecil di dalam bukaan pori telah dieliminasi.

Operasi penutupan, yaitu operasi ganda morfologis, diterapkan pada citra output dari operasi pembukaan. Operasi penutupan adalah aplikasi sekuensial dari operasi *dilatation* yang diikuti oleh operasi *erotion* (yaitu *closing*, *dilation*, *erosion*), keduanya menggunakan elemen penataan yang sama. Operasi penutup menghilangkan semua objek latar belakang, dalam hal ini struktur seperti pori-pori, yang sangat kecil dari SE yang digunakan. Oleh karena itu, sebagai hasil dari operasi penutupan, fitur noise yang terlihat seperti bukaan pori-pori kecil yang disematkan pada serat telah dieliminasi. Penjelasan rinci tentang operasi pembukaan dan penutupan dapat ditemukan oleh Aydilek (2000).

Sebagai hasil dari proses penyaringan morfologi ini, bukaan serat dan pori spesimen geotekstil dapat diidentifikasi dengan lebih mudah. Gambar-31 menunjukkan citra geotextile P setelah operasi penyaringan.

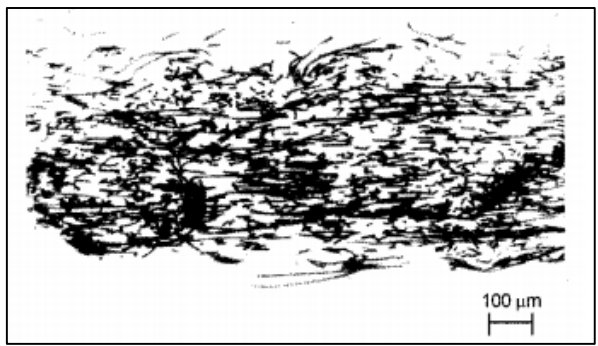

(a)

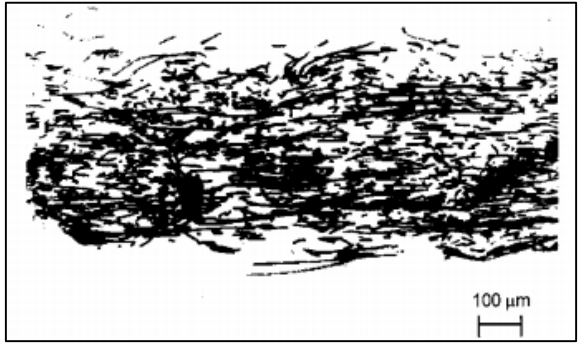

(b)

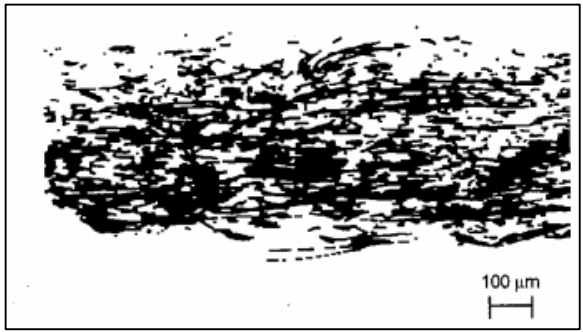

(c)

Gambar-31 Pengolahan citra pada citra penampang lintang geotekstil P, (a) Gambar asli, (b) Gambar biner terukur, (c) Gambar terfragmentasi dan tersaring

Histogram gambar diberi ambang batas pada dua puncak dan pada dua ujung, hitam murni dan putih murni dari sumbu tingkat abu-abu (yaitu pada skala nilai tingkat abu-abu yang berkisar pada 0 dan 255). Verifikasi lebih lanjut dari metode thresholding yang dijelaskan dilakukan dengan dua cara. Pertama, teknik thresholding lokal, sebuah metode yang disarankan oleh Jang dkk (1999) telah digunakan. Bagian lokal pada gambar, mewakili keseluruhan gambar, dipilih dan diukur pada nilai ambang batas percobaan, setelah itu porositas 2D ditentukan. Rata-rata porositas lokal dihitung kemudian dibandingkan dengan porositas seluruh gambar yang diukur pada nilai ambang yang berbeda. Percobaan ini diulang sampai dua nilai ambang.

Cara kedua, porositas planar ditentukan secara manual dan dibandingkan dengan porositas berbasis gambar. Untuk tujuan ini, gambar bagian planar diperbesar dan dicetak pada karton yang memiliki garis kotak-kotak 0,75 mm x 0,75 mm. Kotak di dalam area terbuka dihitung, dan porositas 2D ditentukan dengan membagi area yang dihitung menurut luas pengukuran. Proses ini mirip dengan persen penentuan luas area terbuka dari geotekstil tenun, dan jika dilakukan dengan hati-hati, maka metode tersebut dapat diulang. Seperti yang terlihat dari Gambar-32, porositas gambar yang diukur dengan menggunakan algoritma yang diusulkan sebanding dengan yang ditentukan secara manual dan juga yang diukur dengan teknik *thresholding* lokal.

Algoritma *thresholding* yang dijelaskan di atas digunakan untuk gambar semua geotekstil. Pengecualian adalah geotextile N, yang membutuhkan algoritma khusus. Geotekstil ini adalah geotekstil *nonwoven thermal-bond* dan tidak seperti yang lainnya, seratnya yang relatif tipis menimbulkan masalah pada proses *thresholding* (Gambar-33). Operasi pendeteksian tepi diperlukan untuk mengidentifikasi serat secara jelas sebelum *thresholding*.

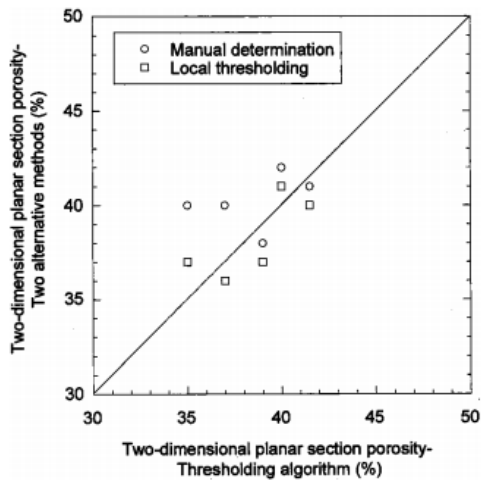

Gambar-32 Verifikasi algoritma thresholding dengan metode alternatif

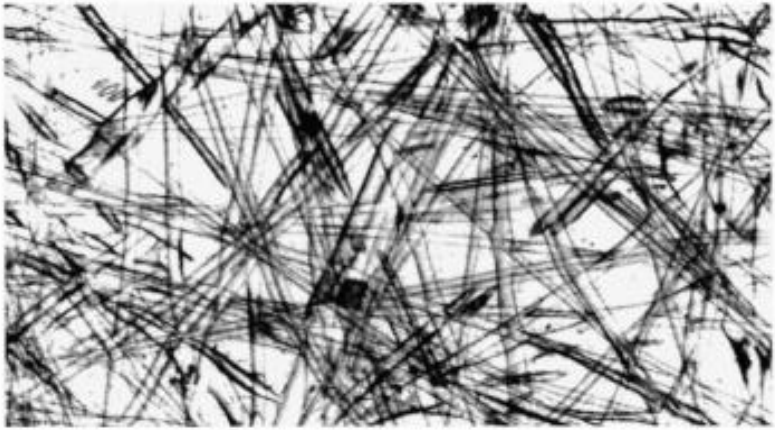

(a)

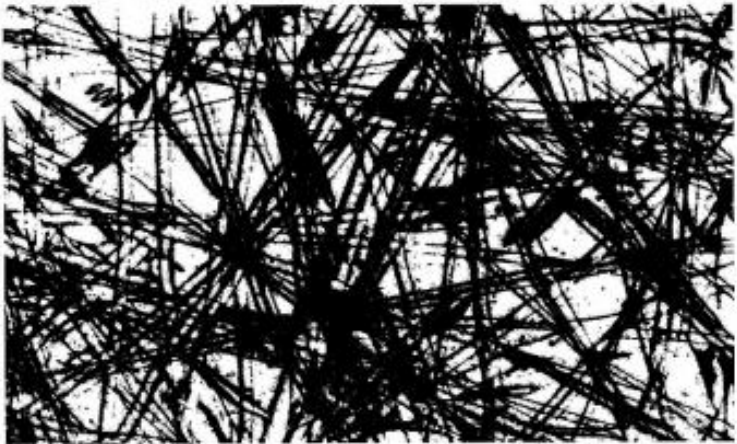

(b)

Gambar-33 Gambar planar dari geotekstil N, (a) Gambar asli, (b) Citra yang kurang jelas disebabkan oleh masalah tepi

Model ini terutama didasarkan pada analisis gambar penampang melintang. Jika porositas dan PSD berdasarkan gambar planar hampir sama sepanjang seluruh ketebalan geotekstil, maka irisan penampang melintang dapat digunakan untuk menyelidiki kontinuitas berbagai bukaan pori-pori melalui ketebalan geotekstil. Dengan cara itu, gambar penampang melintang dapat memberikan informasi yang diperlukan tentang struktur pori 3D. Untuk mendemonstrasikan ini, porositas dan PSD irisan surfaktan paralel dari geotekstil ditentukan dengan menggunakan algoritma yang dikembangkan di Imaq, perangkat lunak analisis citra yang bekerja di bawah perangkat lunak LabVIEW. Algoritma ini didasarkan pada metode ekivalensi bentuk, dan deskripsi singkat algoritma diberikan di bagian selanjutnya. Gambar-33 diberikan sebagai contoh proses yang dilakukan pada geotekstil P. Oleh karena itu, disimpulkan bahwa gambar penampang melintang akan mewakili struktur pori geotekstil, dan digunakan untuk analisis ukuran pembukaan pori-pori. Masounave dkk (1980) melaporkan temuan konsisten yang membuktikan bahwa gambar penampang melintang dapat mewakili PSD medium geotekstil dengan cukup baik.

Untuk memberikan informasi tentang distribusi ukuran pori pembukaan gambar cross-sectional, diperlukan operasi pengiris. Sebuah algoritma baru dikembangkan untuk mengiris gambar cross-sectional secara optimal.

Jumlah irisan ditentukan dengan membagi ketebalan geotekstil dengan ketebalan serat rata-rata. Prosedur iteratif yang dijelaskan di sini memposisikan irisan sedemikian rupa sehingga menghasilkan *slicing grid* yang paling sesuai dengan segmen serat yang berorientasi horizontal (yang diamati pada citra penampang melintang). Setelah *grid* pengiris diposisikan secara optimal, porositas 2D masing-masing irisan dihitung. Porositas irisan 2D ini ditentukan sepanjang potongan, maka dari itu variabel ini diberi nama '' porositas longitudinal.

Diagram alir algoritma pengiris dapat dilihat pada Gambar-34. Gambar-35 (a-h) mengilustrasikan langkah-langkah algoritma pengiris yang diterapkan pada citra penampang lintang geotextile P. Untuk kejernihan visual, segmen gambar diperbesar dan ditunjukkan pada gambar-gambar ini. Ada dua tahap dasar algoritma pengiris:

• penentuan ketebalan serat rata-rata pada resolusi pengambilan gambar penampang melintang dan pembangkitan grid pengiris seragam; dan

• penentuan posisi optimal dari grid pengiris seragam pada citra penampang melintang.

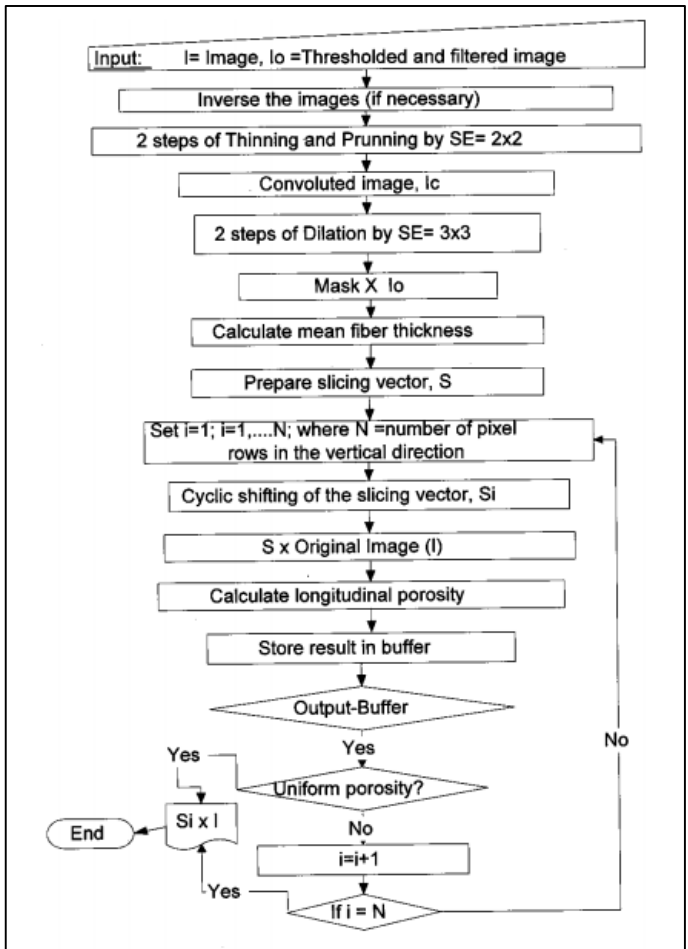

Gambar-34 Diagram alir dari algoritma pengiris

Dalam gambar biner, secara *default*, *pixel* yang terkait dengan daerah serat dan pori memiliki nilai 0 (berwarna hitam) dan 1 (berwarna putih). Karena serat geotekstil harus diproses dengan algoritma pengiris untuk penentuan ketebalan serat, pelengkap gambar biner ini diambil. Gambar 10 (a) menunjukkan segmen gambar penampang melintang dengan warna terbalik (warna putih menjadi hitam dan hitam menjadi putih) dari geotextile P.

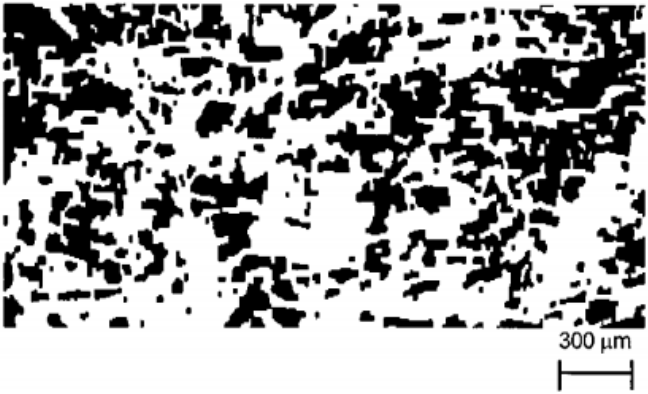

(a)

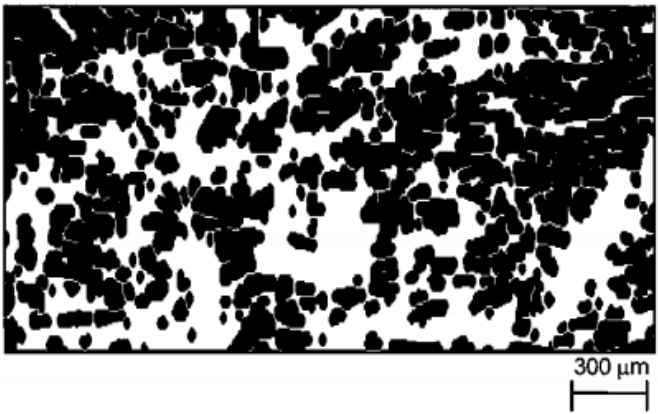

(b)

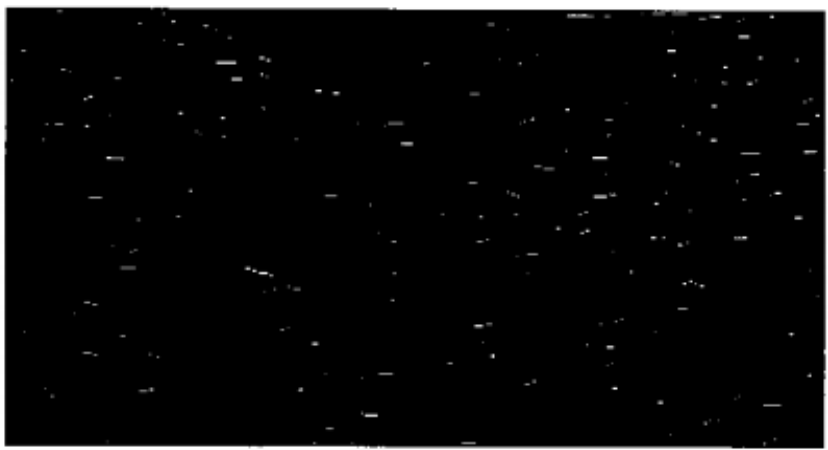

(c)

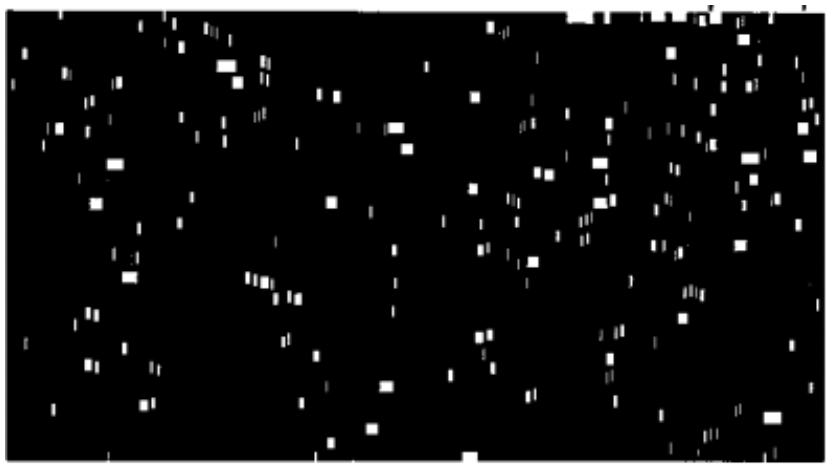

(d)

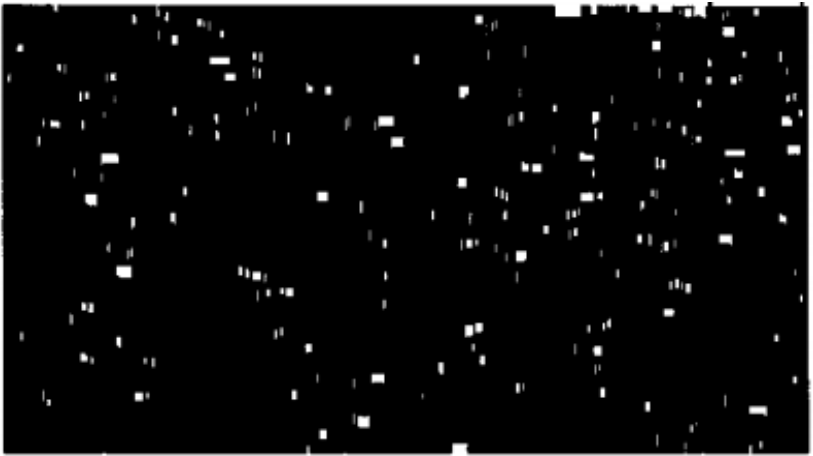

(e)

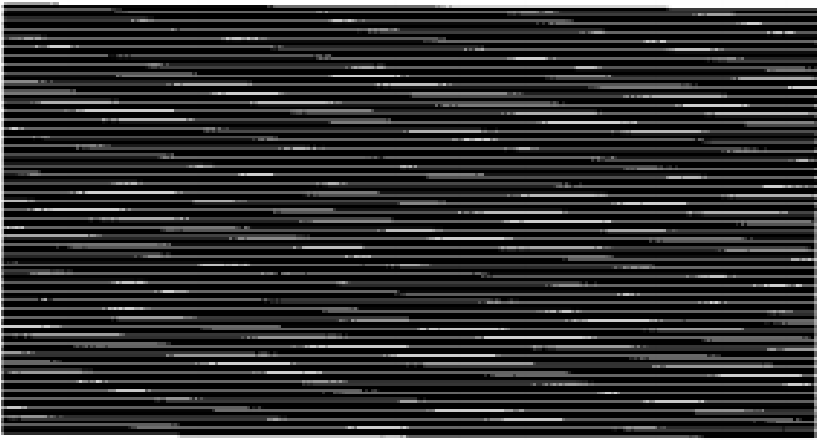

(f)

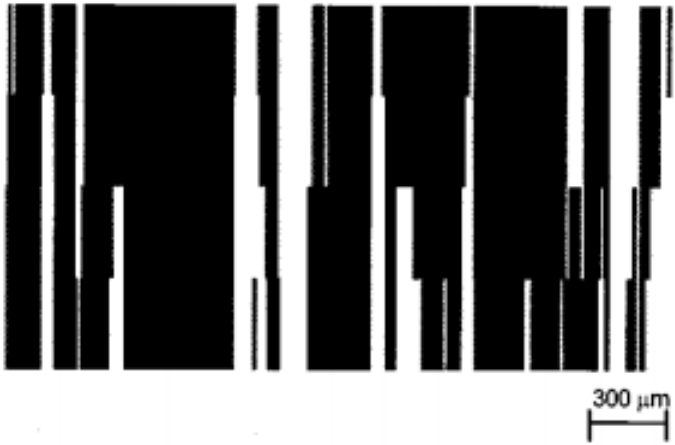

(g)

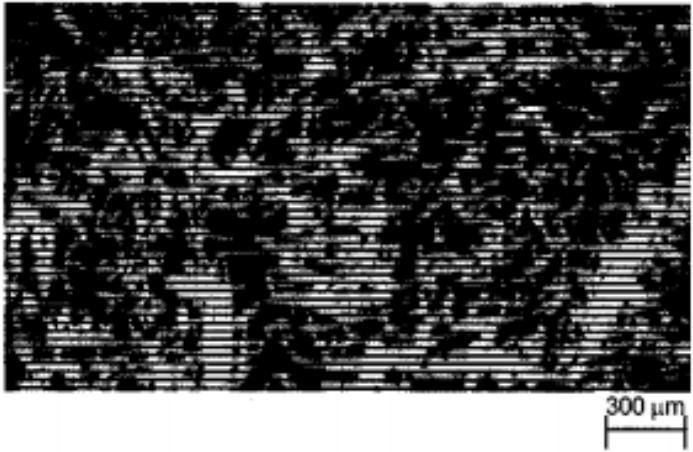

(h)

Gambar-35 Gambar planar geotekstil P: (a) Gambar terfragmentasi dan tersaring; (b) Penerapan operasi penipisan dan pemangkasan; (c) Operasi konvolusi untuk mengekstrak serat horizontal; (d) Proses pelebaran (dilatasi); (e) Masking dari gambar asli; (f) *Optimal slicing vector* yang digunakan dalam proses mengiris; (g) Gambar selama pergeseran siklik; (h) Hasil dari s*licing vector* yang dikalikan dengan gambar aslinya

Berdasarkan hasil pengukuran porositas menggunakan metoda tersebut, didapatkan hasil seperti pada Gambar-36. Ilustrasi kenampakan kain yang diolah oleh algoritma dapat dilihat pada Gambar-37.

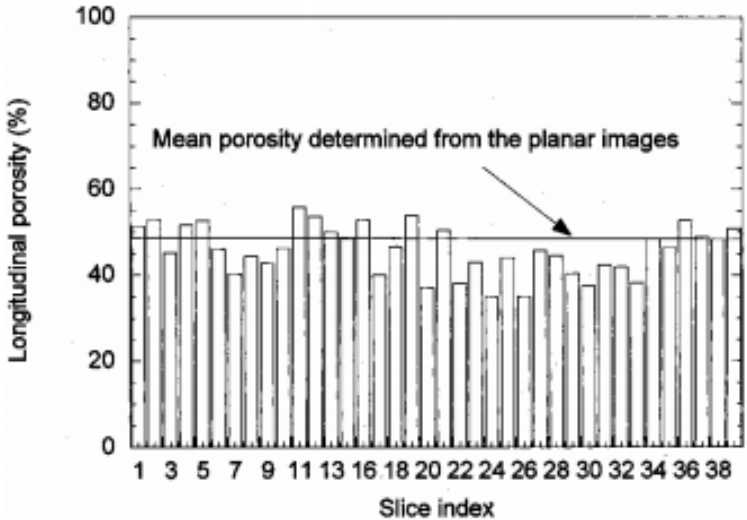

Gambar-36 Hasil analisis porositas penampang melintang rata-rata kain geotekstil

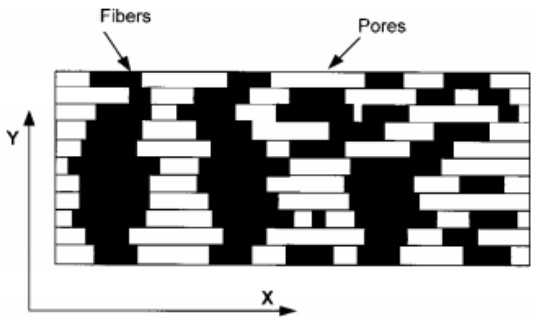

Gambar-37 Ilustrasi kenampakan penampang melintang geotekstil yang dikenali oleh algoritma komputer

Dapat dilihat, bahwa hasil pengukuran porositas kain berdasarkan pengukuran planar dan pengukuran pada penampang melintangnya dapat memberikan nilai yang berbeda dan bervariasi. Hal tersebut dikarenakan pengukuran pada bidang planar kain geotekstil belum dapat menterjemahkan kain sebagai struktur 3D, melainkan hanya sebagai struktur 2D.

Setelah pengenalan serat dan pori-pori pada irisan, selanjutnya besarnya nilai *pore opening size* diukur dengan menggunakan suatu algoritma yang dikembangkan pada perangkan lunak Imaq. Imaq memiliki suatu metode pengukuran yang dapat mengukur objek dengan suatu algoritma tertentu. Distribusi nilai varians dan nilai besarnya bukaan pori pada geotekstil dapat dilihat pada Gambar-38.

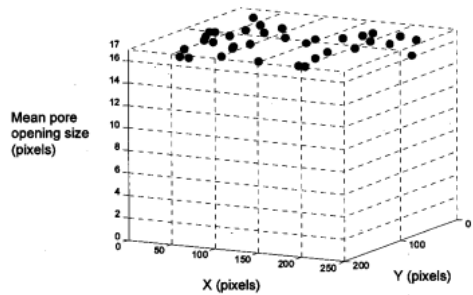

(a)

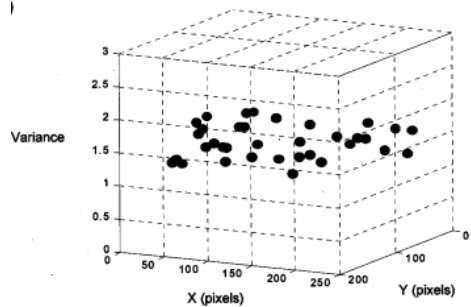

(b)

Gambar- 38 Distribusi nilai varians dan nilai besarnya bukaan pori geotekstil

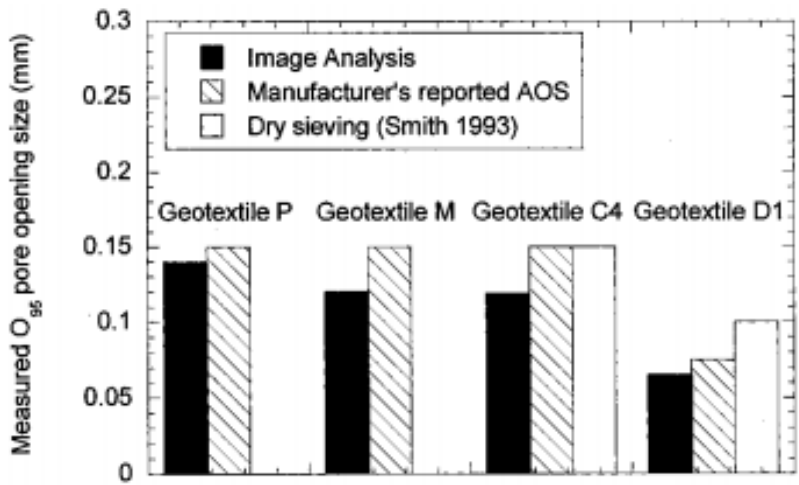

(a)

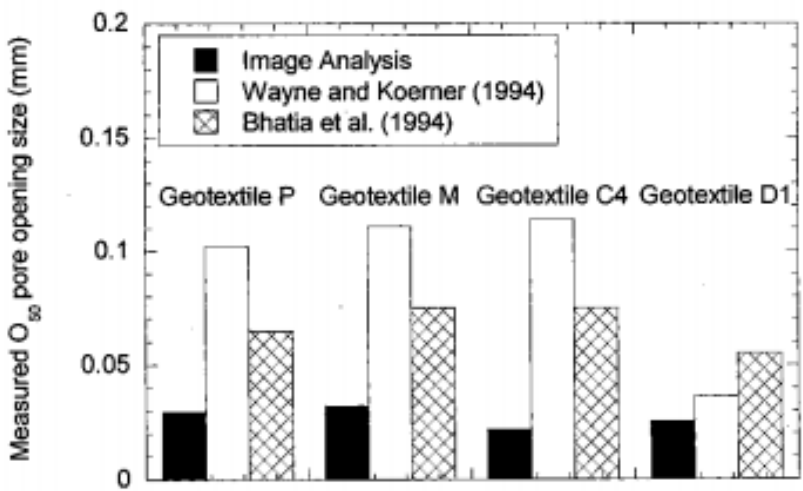

(b)

Gambar-39 Perbandingan hasil analisis besar pori kain geotekstil *nonwoven image processing* pada berbagai metode uji yang telah ditemukan

Berdasarkan perbandingan yang ada pada grafik Gambar-39, ditemukan bahwa metoda pengukuran pori $O_{95}$ dan $O_{50}$ *image processing* akan menghasilkan nilai pengukuran yang relatif lebih kecil dibandingkan dengan nilai hasil pengukuran metode lainnya. Hal tersebut berlaku bagi kain geotekstil P, geotekstil M, geotekstil C4 dan geotekstil D1.

Academy of Science, Engineering & Technology;Sep2009, Issue 33, p543.

[82] Serdaroglu A et al (2004). Defect detection in textile fabric image using wavelet transforms and independent component analysis, Pattern Recognition and Image Understanding, 18-23.

[83] Sezer O G et al (2003). Independent component analysis for texture defect detection, the 6th German-Russian workshop of Pattern Recognition and Image Understanding, Katun Village, Novosibirsk, 210-213.

[84] Smith, J. L. (1993). ''The pore size distribution of geotextiles.'' MS thesis, Syracuse Univ., Syracuse, N.Y.

[85] Sonka, M., Hlavac, V. and Boyle, R. Image Processing, Analysis, and Machine Vision, 1999 (PWS Publishing, PaciŽ c Grove, California; London).

[86] Stojanovic R et al (1999). Automated Detection And Neural Classification Of Local Defects in textile web, Seventh International Conference on image processing-IPA99, Manchester, UK

[87] T. J. Kang, S. C. Kim, I. H. Sul, J. R. Youn, and K. Chung. 2005.Fabric Surface Roughness Evaluation Using Wavelet-Fractal Method Part I: Wrinkle, Smoothness and Seam Pucker, Textile Res. J. 75(11), 751–760.

[88] The MathWorks Inc., Matlab Wavelet Toolbox Version 3.0: The MathWorks Inc., 2004

[89] The Woolmark Company, "Pilling a curse of the knitwear industry," News, 2000; http://melpub.wool.com/enews2.nsf/vwMonthlyWoolmark/e7633a84 92 4786304a25695e00797adc?OpenDocument&Archive.

[90] V. Sülar, and A. Okur. 2008. Objective Evaluation of Fabric Handle by Simple Measurement Methods, Textile Res. J. Vol 78(10): 856–868.

[91] Wayne, M. H., and Koerner, R. M. (1994). ''Correlation between longterm flow testing and current geotextile filtration design practice.'' Proc., Geosynthetics '93, Industrial Fabrics Association International, Roseville, Minn. 501–517.